\documentclass[aps,prx,noshowpacs,superscriptaddress,float-fix,12pt,final,times]{revtex4-2}
\usepackage{etoolbox} 
\usepackage{lipsum} 
\usepackage[colorlinks,citecolor=blue,pdftex]{hyperref}
\usepackage{graphicx}
\usepackage{dcolumn}
\usepackage{bm}
\usepackage[pdftex]{hyperref}
\usepackage{amssymb} 
\usepackage{amsmath}
\usepackage{wrapfig}
\usepackage{float}
\usepackage{color}
\usepackage{times}
\usepackage{textgreek}
\usepackage{hyperref}
\usepackage{xcolor}

\begin{document}
	
\title{Underdamped harmonic oscillator driven by a train of short pulses: Analytical analysis}

  \author{Chanseul Lee}%
\affiliation{%
	Department of Physics, Korea University, 145 Anam-ro, Seongbuk-gu, Seoul 02841, Republic of Korea}
\author{Tai Hyun Yoon}%
  \email{thyoon@korea.ac.kr (THY)}
  \affiliation{%
  	Center for Molecular Spectroscopy and Dynamics, Institute for Basic Science (IBS), Korea University, 145 Anam-ro, Seongbuk-gu, Seoul 02841, Republic of Korea}
  \affiliation{%
  	Department of Physics, Korea University, 145 Anam-ro, Seongbuk-gu, Seoul 02841, Republic of Korea}

\date{\today}



\begin{abstract}
A theoretical model of an underdamped harmonic oscillator (UHO) driven by periodic short pulses may find plenty of applications in classical, semiclassical, and quantum physics. We present here two different forms of analytical solutions: {\it time-periodic solutions} and {\it harmonic solutions} for one-dimensional classical UHO driven by three different trains of short pulses. They are a Dirac comb, a train of square pulses, and a train of Gaussian pulses with the same pulse-to-pulse time interval $T$ and pulse width $2\tau$. Two solutions for square and Gaussian pulses approach to that of the Dirac comb when the pulse width $2\tau \rightarrow 0$ as expected. In particular, the harmonic solutions for Dirac comb and Gaussian pulses could be expressed approximately with harmonic terms of the repetition frequency $\omega_{\rm R} = 2\pi/T$ up to the second order. The presented analytical solutions would provide a practical way to determine experimentally the system parameters such as the underdamped oscillation frequency $\omega = \sqrt{\omega_0^2-\gamma^2}$, the natural frequency $\omega_0$, and the damping rate $\gamma$, by nonlinear curve fitting procedures for different driving force parameters of $T$ and $2\tau$. 
\end{abstract}

\maketitle



\section{Introduction \label{sec:Intra}}

Periodically driven harmonic oscillator systems are of paramount importance in the classical \cite{Yeon87,Kells04,Tayler05,Choi08,Tang08,Marion09,Gao13,Zech21} and quantum sciences \cite{Gzyl83,Um87,Um02,Mudde03,Goldman15,Prado17} and thus can find plenteous applications across the whole scientific and technological disciplines. In particular, a model of damped harmonic oscillator \cite{Dekker81} driven by a train of delta-kick forces \cite{Tayler05,Mudde03,Kells04} has been provided theoretical understandings of the temporal behaviors of the physical oscillators under the influence of periodic short pulses. Among various damped oscillator models \cite{Dekker81}, the underdamped harmonic oscillator (UHO) \cite{Goldstein02,Tayler05,Marion09}, where underdamped oscillation frequency $\omega = \sqrt{\omega_0^2 - \gamma^2}$ is positive and smaller than the natural frequency $\omega_0$ and the system decay rate $\gamma$, has been attracted much attentions \cite{Goldstein02} since it exhibits multi-phase temporal behaviors depending on the relative time scales between the ststem and driving force parameters. Thus, precise {\it a priori} knowledge of the system parameters of the UHO driven by a triain of short pulses  is of atmost importance before any experimental and theoretical studies. 

Specifically, the temporal behavior of the driven UHO is determined by the relative magnitudes between the system parameters of the UHO  (oscillation period $t_0 = 1/\omega$ and damping time $\tau_0 = 1/\gamma$) and the parameters of the driving forces (pulse width $2\tau$ and pulse-to-pulse time interval $T$) \cite{Wright04,Gao13,Gao14,Gao15,Hassan20}. Here, in this paper, we present general analytical solutions of the UHO driven by a train of three different trains of short driving forces $(T > 2\tau)$, i.e., a Dirac delta comb, a train of square pulses, and a train of Gaussian pulses. The first two models of periodic driving forces provide analytical solutions particularly useful for theoretical and numerical understanding of the temporal befaviors of the various UHOs under the influence of periodic perturbations \cite{Tayler05,Goldstein02}. In addition, the train of Gaussian pulse model \cite{Wright04,Gao13,Gao14,Gao15,Hassan20,Alharbey21} provides analytical solutions to be directly applicable to understand the temporal behaviors of the UHOs to determine the system parameters $\omega, \,\omega_0$, and $\gamma$ experimentally by fitting the analytical solutions to the experimental data for different values of driving force parameters $2\tau$ and $T$. We confirm that the analytical solutions for the trains of square and Gaussian pulses become identical to the Dirac comb solutions when the pulse width approaches to zero, i.e., $2\tau \rightarrow 0$.

\begin{figure}[tb]
	\includegraphics[scale=0.6]{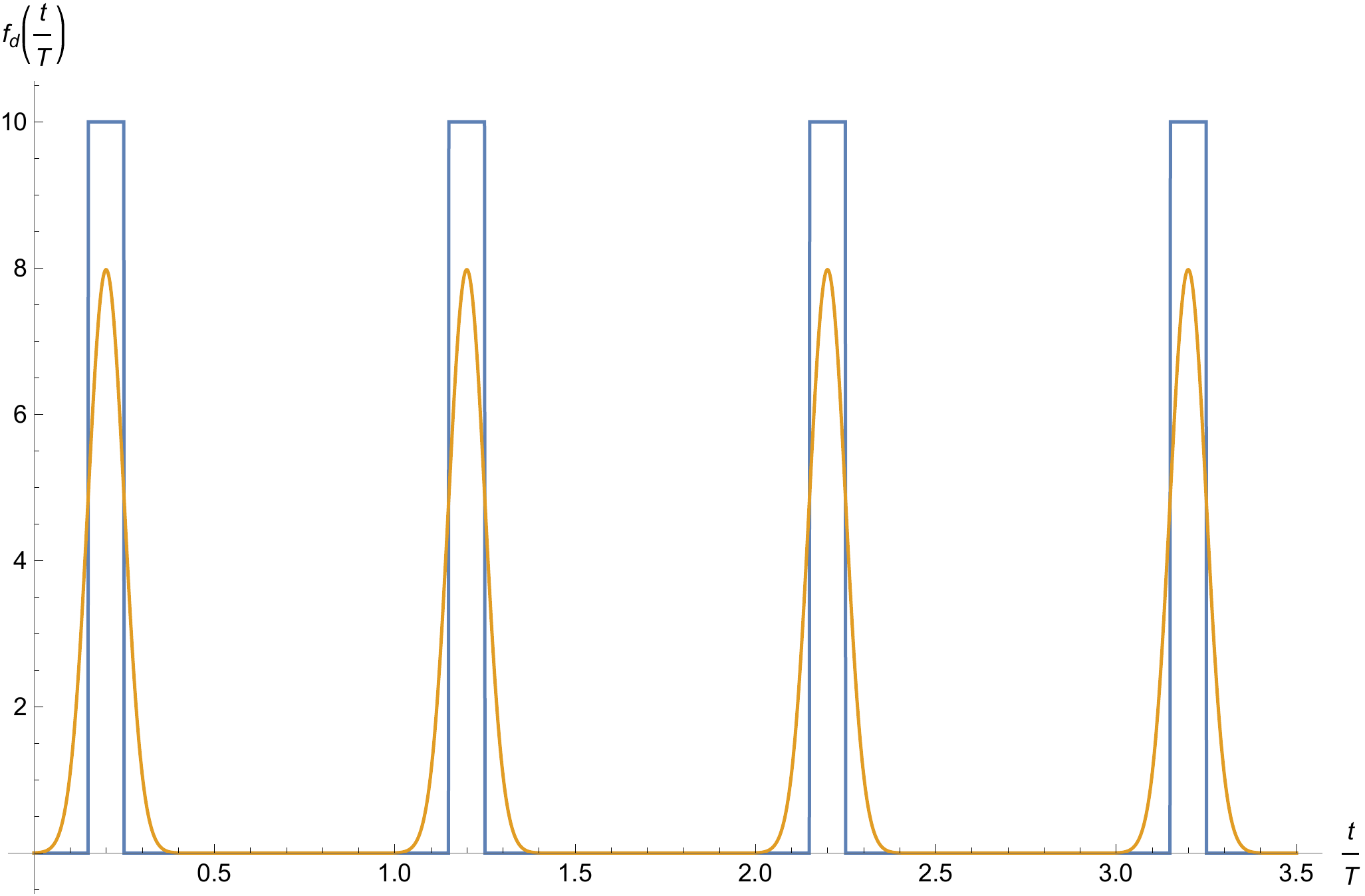}
	\caption{First four short pulses of two different normalized periodic driving forces, $f_d(t/T) = F_d(t,\tau,T)/F_0$, where $F_0$ is the peak force, i.e., the square pulses (blue line) and the Gaussian pulses (orange line), with the same single-pulse areas with the parameters of $2\tau/T = 0.1$ and $Q/T= 0.2$.}\label{Fig1}
\end{figure}

To find the analytical solutions, we consider first typical theoretical models of the periodic pulse train \cite{Jones00,Hassan20}, $F_d(t,\tau,T)$, such as a Dirac delta comb, which may be represented by taking a zero-width limit ($\tau\rightarrow0$) of the train of square pulses that has a unit pulse area and the same pulse-to-pulse interval $T$. Although two models, i.e., the Dirac delta comb and the train of square pulses, provide a meaningful road to simulate theoretical responces of the UHO, they are far from real interaction models of the driving forces, since the Dirac comb model has a nonrealistic zero-pulse width and the square pulse model has flat interaction strength within the finite pulse width $2\tau$ (see Fig. \ref{Fig1}). On the other hand, a theoretical model of periodic Gaussian pulses \cite{Wright04,Gao13,Gao14,Gao15,Hassan20,Alharbey21} that has a unit pulse area and pulse width of $2\sqrt{2}\tau$ may be a realistic (experimentally realizable) delta-function representation at the limit of untra-short pulse width (2$\tau\rightarrow0$). Figure~\ref{Fig1} compares, for example, the first four pulses of two different normalized periodic driving forces $f_d(t/T) = F_d(t,\tau,T)/F_0$, where $F_0$ is the peak force, i.e., the square pulses (blue line) and Gaussian pulses (orange line), with the same single-pulse areas of one with the parameters of $\tau/T = 0.05$ and $Q/T= 0.2$, where $Q$ is the time shift parameter.

In this paper, we present two sets of complete analytical solutions of the UHO \cite{Dekker81} driven by three different trains of short driving forces: One set of {\it time-periodic solutions} in Sec.~\ref{sec:ATS} and the other set of {\it harmonic solutions} in Sec.~\ref{sec:AHS}. The presented analytical solutions would provide, we believe, the coprehensive understandings and comparisons of temporal behaviors of the three different driving force models depending on the time scales of UHO system parameters, i.e., $\tau_0$ and $t_0$ versus the parameters of the driving forcees, i.e., $2\tau$ and $T$. Furthermore, the presented analytical solutions would be useful to determine the system parameters of the UHOs $\omega, \,\omega_0$, and $\gamma$ experimentally by nonlinear curve fitting the analytical solutions to the experimental data at different regimes of driving force parameters $2\tau$ and $T$.

\section{UHO driven by a train of short pulses  \label{sec:System}}

In this section, we consider an UHO in one-dimension \cite{Dekker81} driven by a train of periodic forces, $F_{\rm d}(t,\tau,T)$, consisting of $(N_d+1)$ identical pulses with the pulse width of 2$\tau$ and pulse-to-pulse interval (period) of $T$, i.e., with the pulse repetition rate $\omega_{\rm R} =2\pi/T$. Then, the driving force can be written as 
\begin{equation}\label{PForce}
F_{\rm d}(t,\tau,T) = \sum_{n=0}^{N_{\rm d}} F_d(t-nT,\tau).
\end{equation}
The equation of motion of the driven DHO can then be usually written as  
\begin{equation}\label{EOM}
	\ddot{x}(t) + 2\gamma \dot{x}(t) + \omega_0^2 x(t) =\frac{1}{m}F_{\rm d}(t,\tau,T),
\end{equation}
where $x(t)$ is the displacement of the DHO  at time $t$, dot means a time derivative, $m$ being the mass of the particle, $\gamma = b/2m$ being the damping rate, $b$ being the damping coefficient, $\omega_0=\sqrt{k/m}$ being the natural resonance frequency, and $k$ being the restoring force constant. 

The solution of Eq.~(\ref{EOM}) can be obtained formally by using the Laplace transforms \cite{Afken13} such that
\begin{equation}\label{solx}
	x(t) = x_{\rm h}(t) + x_{\rm p}(t,\tau,T),
\end{equation}
where $x_{h}(t)$ is the homogeneous solution obtained by assuming $F_{\rm d}(t,\tau,T) = 0$ and $x_{\rm p}(t,\tau,T)$ is the particular solution. From Eq.~(\ref{EOM}), the homogeneous solution $x_{\rm h}(t)$ can be easily obtained by taking Laplace transform of $x(t)$, $\mathcal{L}_{\rm h}(s) = \mathcal{L}[x(t)]$,  into the complex $s$ domain with the initial conditions $x(0) = x_0$ and $\dot{x}(0) = v_0$ as the follow
\begin{equation}\label{Lsolxh}
	\mathcal{L}_{\rm h}(s) = \frac{(s+2\gamma)x_0 + v_0}{s^2+2\gamma s + \omega_0^2} = \mathcal{H}(s)\left((s+2\gamma)x_0 + v_0\right),
\end{equation}
where 
\begin{equation}\label{TRF}
\mathcal{H}(s) = \frac{1}{s^2+2\gamma s + \omega_0^2} = \frac{1}{(s-s_+)(s-s_-)},
\end{equation}
and $\mathcal{H}(s)$ is the transfer function of the DHO, $s_\pm = -\gamma \pm i\omega$, $\omega = \sqrt{\omega_0^2-\gamma^2}$ is the underdamped oscillation frequency at the weak damping limit. Then, the condition $0 < \gamma < \omega_0$ holds for any UHO as in the case of the present work. We see that the inverse Laplace transform   $h(t)$ of $\mathcal{H}(s)$ in Eq.~(\ref{TRF}) is simply given by 
\begin{equation}\label{ILTFGs}
	h(t) = \mathcal{L}^{-1}[\mathcal{H}(s)] = \frac{1}{\omega}e^{-\gamma t}\sin(\omega t).
\end{equation}

Now, the homogeneous solution $x_{\rm h}(t)$ in the time domain in Eq.~(\ref{solx}) can be obtained by taking the inverse Laplace transform  (or by the Bromwich integral \cite{Afken13}) of Eq.~(\ref{Lsolxh}) as the follow
\begin{equation}\label{ILhxt}
	x_{\rm h}(t) = \mathcal{L}^{-1}[\mathcal{L}_{\rm h}(s)] = \frac{1}{2\pi i}\int_{\ell-i\infty}^{\ell+i\infty}  e^{st}\frac{(s+2\gamma)x_0+v_0}{(s-s_+)(s-s_-)} ds, 
\end{equation}
where $\ell \in \mathbb{R}$ that exceeds the real parts of all the singularities in Eq.~(\ref{ILhxt}). The integrand $I_{\rm h}(s,t)$ of Eq.~(\ref{ILhxt}) has two simple poles, $s = s_{\pm}$, at the roots of the quadratic equation in the denominator. Since the two poles $s_{\pm}$ lie to the left of $s = 0$ in the complex plane for $\gamma > 0$, we may perform the integration in Eq.~(\ref{ILhxt}) along the imaginary axis and close in the left-half plane, where the exponential sends the integrand $I_{\rm h}(s,t)$ to zero. By the residue theorem \cite{Afken13}, the integral $ I_{\rm h}(t) = \int_{\ell-i\infty}^{\ell+i\infty} I_{\rm h}(s,t) ds$ is $2\pi i $ times the sum of two residuals, i.e.,
\begin{eqnarray}\label{ILsolxh}
	x_{\rm h}(t) &=& {\rm Res}[I_{\rm h}(s,t)]_{s_+}+{\rm Res}[I_{\rm h}(s,t)]_{s_-} \nonumber \\
	&=& e^{s_+t}\frac{(s_++2\gamma)x_0+v_0}{s_+-s_-} + e^{s_-t}\frac{(s_-+2\gamma)x_0+v_0}{s_{-}-s_+} \nonumber \\
	&=&  e^{-\gamma t}\left(x_0\cos(\omega t) + \frac{v_0+\gamma x_0}{\omega}\sin(\omega t)\right).
\end{eqnarray}
From Eq.~(\ref{ILsolxh}), we see that the inverse Laplace transform $h(t) = \mathcal{L}^{-1}[\mathcal{H}(s)]$ in Eq.~(\ref{ILTFGs}) is nothing but the homopgeneous solution $x_{\rm h}(t)$ with the initial conditions $x_0 = 0$ and unit inital velocity initiated by the unit impulse force, i.e., Green's function \cite{Jackson99}. Note also that for $t>\tau_0 = 1/\gamma$, the homogeneous solution $x_{\rm h}(t)$ in Eq.~(\ref{ILsolxh}) damped out completely after a single delta-kick driving at $t = 0$. Thus, if the pulse-to-pulse interval $T$ is shorter than the system decay time $\tau_0$, i.e., $T < \tau_0$, the temporal behavior of the UHO could exibit a temporal interference between the finite number of pulses witin $t < \tau_0$ \cite{Wright04,Gao13,Gao14,Gao15,Hassan20}.

\section{Particular solutions \label{sec:PS}}

As briefly mentioned in the previous section, when an UHO is driven by the periodic forces  $F_{\rm d}(t,\tau,T)$ in Eq.~(\ref{PForce}), one needs to consider the relation between two characteris times between the system decay time $\tau_0 = 1/\gamma$ and the pulse-to-pulse time interval $T > 2\tau$. Since $\omega_0 > \gamma$ in the UHO, the temtoral behavior is governed not by the higher natural frequency $\omega_0$, but by the lower system damping frequency $\gamma$. In other words, the oscillation period of the harmonic oscillation $t_0 = 1/\omega_0$ is shorter than $\tau_0$, i.e., $\tau_0 > t_0$ in the weak damping limit. If $\tau_0 < T$ , the homogeneous solution $x_{\rm h}(t)$ is nonzero only in the time interval between $0 < t < T$. On the other hand, for the UHO at the intermediate ($\tau_0 \approx T$) or at the long system decay time ($\tau_0 > T$), the homogeneous solution $x_{\rm h}(t)$ can survive over many periodic pulses so that it should be overlapped with the particular solutions.
	
After $t \gg \tau_0$, i.e., the UHO exhibits an assymptoic behavior by the particular solution $x_{\rm p}(t)$ only. The particular solution of Eq.(\ref{solx}) that depends now on the periodic driving force $F_{\rm d}(t,\tau,T)$ can be obtained by two-step processes as same as the case obtaining the homogeneous solution above: In the first step, by taking the Laplace transform of Eq.~(\ref{EOM}), we obtain $\mathcal{L}_{\rm p}(s)$ such that 
\begin{equation}\label{Lsolxp}
	\mathcal{L}_{\rm p}(s) = \mathcal{H}(s)\mathcal{F}_{\rm d}(s,\tau,T),
\end{equation}
where $\mathcal{F}_{\rm d}(s,\tau,T) = \mathcal{L}\left[F_{\rm d}(t,\tau,T)\right]$, and then we take, in the second step, the inverse Laplace transform of Eq.~(\ref{Lsolxp}) to obtain $x_{\rm p}(t)$ as the follow
\begin{equation}\label{ILsolxp}
	x_{\rm p}(t,\tau, T) = \mathcal{L}^{-1}\left[ \mathcal{H}(s) \mathcal{F}_{\rm d}(s,\tau,T)\right].
\end{equation}

To find the particular solution $x_{\rm p}(t,\tau,T)$ in Eqs.~(\ref{solx}) and (\ref{ILsolxp}), we need to take into account the specific temporal form of the periodic driving force $F_{\rm d}(t,\tau,T)$ in Eq.~(\ref{EOM}) as introduced in the previous section. 
In order to compare the particular solutions, $x_{\rm p}(t)$s in Eq.~(\ref{ILsolxp}), corresponding to three different periodic driving forces, we introduce a time shift parameter $Q>0$ to match the center of each pulses at $t = nT + Q$ with the pulse number $n$ for the train of short pulses, i.e., $T > Q, \tau$, and $Q > \tau$ (see Fig.~\ref{Fig1}). Then, the periodic driving force in Eq.~(\ref{PForce}) can now be written as 
\begin{equation}\label{FQd}
	F_{\rm d}^{\rm K}(t,\tau,T,Q) = \sum_{n=0}^{N_{\rm d}} F_{\rm d}^{\rm K}(t -nT-Q,\tau),
\end{equation}
where ${\rm K} = {\rm DC}, {\rm SP}$ and ${\rm GP}$ indicate, respectively, the Dirac comb, the square pulses, and the Gaussian pulses. 

To perform the inverse Laplace transform in Eq.~(\ref{ILsolxp}), we try two different approches: First, a {\it time-periodic solutions}, in which we take the inverse Laplace transform for the individual pulse $n$, then sum over all particular solutions for each pulses $n$ to get the particular solutions, i.e., $x_{\rm TP}^{\rm K}(t) = \sum_{n=0}^{N_d}\mathcal{L}^{-1}[\mathcal{H}(s)F_{\rm p}^{\rm K}(s,n)]$, where $F_{\rm p}^{\rm K}(s,n)$ is the Laplace transform of $F_{\rm d}^{\rm K}(t -nT-Q,\tau)$. Second, a {\it harmonic solutions}, in which we sum over all pulses with the limit $N_d\rightarrow\infty$, then take the inverse Laplace transform at once to get  $x_{\rm HS}^{\rm K}(t) = \mathcal{L}^{-1}[\mathcal{H}(s)\sum_{n=0}^\infty F_{\rm p}^{\rm K}(s,n)]$. The time-periodic solutions for three different driving forces are obtained in the Appenx \ref{sec:ATS}, while the corresponding harmoic solutions are obtained in the Appendix \ref{sec:AHS}.

As we shall see in the next section, the former method (time-periodic solutions) results in simple and informative expressions of the particular solutions. On the other hand, the later method (harmonic solutions) results in the particular solutions as the closed froms of an infinite sums of discrete Fourier frequencty components at $\omega_k = k \omega_{\rm R} =  2\pi k/T, k = 0,1,2, \cdots$ \cite{Tayler05}. Both analytical results show exactly the same results in the numerical solutions with the same system and driving force parameters accorss large range of parameter values, but manifest themselfs differently in analytical expressions depending on the ranges of temporal parameters of the system and driving forces. 

\section{Analytical Analysis and Discussions \label{sec:Analysis}}

In the Appendix \ref{sec:ATS} and Appendix \ref{sec:AHS}, we found two sets of analytical solutions for an UHO driven by a train of short pulses, i.e., the time-periodic solutions in Sec.~\ref{sec:ATS} and the harmonic solutions in Sec.~\ref{sec:AHS}, for three different periodic driving forces. The potential applications of the analytic solutions presented in this paper would be far from current expectations, since the analytical solutions are valid for vast ranges of system and driving force parameters. In this section, we explore a few basic temporal behaviors of the different solutions at different time scales between the time parameters of the driving forces, i.e., $2\tau$ (pulse width) and $T$ (pulse-to-pulse time interval) compared to the system time parameters, i.e., $t_0 = 1/\omega$ (oscillation period of the UHO) and $\tau_0 = 1/\gamma$ (system energy daming time).

The general solution of the equation of motion in Eq.~(\ref{EOM}) for the UHO driven by a train of periodic forces $F_d(t,\tau,T)$ is given by 
\begin{equation}\label{Fsol}
	x^{\rm K}(t) = x_{\rm h}(t) + x_{\rm p}^{\rm K}(t,\tau,T),
\end{equation}
as discussed in Eq.~(\ref{solx}), where ${\rm K = DC, SP, GP}$, respectively, stand for the train of the Dirac comb, square pulses, and Gaussian pulses.  The homogeneous solution $x_{\rm h}(t)$ in Eq.~(\ref{Fsol}) is the same for all $x^{\rm K}(t)$ as given in Eq.~(\ref{ILsolxh})
\begin{equation}\label{Hosol}
x_{\rm h}(t) =  e^{-\gamma t}\left(x_0\cos(\omega t) + \frac{v_0+\gamma x_0}{\omega}\sin(\omega t)\right).
\end{equation}

From now on, we consider only the case when the system starts to oscillate at its equilibrium position and driven by the periodic pulses $f_d(t,\tau,T)$ at $t = 0$ with initial conditions $x_0 = 0$ and $v_0 = 0$. Then, the homogeneous solution does not contribute at all for $x^{\rm K}(t)$ since $x_{\rm h}(0) = 0$. 
We now focus on the particular solutions $x^{\rm K}(t) = x_{\rm p}^{\rm K}(t,\tau,T)$ for $t\ge 0$.

\subsection{Dirac comb driving \label{sec:NCD}}

We found, in Appendix \ref{sec:ATS} and \ref{sec:AHS}, two particular solutions for Dirac comb driving: the time-periodic solution $x_{\rm TP}^{\rm DC}(t)$ given in Eq.~(\ref{XpsolDC}) \cite{Marion09} and the harmonic solution $x_{\rm HS}^{\rm DC}(t)$ given in Eq.~(\ref{xpt}) as follows
\begin{subequations}\label{DDC}
\begin{eqnarray}
	x_{\rm TP}^{\rm DC}(t) 
	&=&\frac{I_p}{m\omega}\sum_{n=0}^{N_d} e^{-\gamma (t-nT-Q)}\sin(\omega(t-nT-Q))\Theta(t-nT-Q), \label{DDCa} \\
	x_{\rm HS}^{\rm DC}(t) &=& 
	\frac{I_pe^{-\gamma (t-Q)}}{m\omega(1-2e^{\gamma T}\cos(\omega T)+e^{2\gamma T})}\left(\sin\left(\omega(t-Q)\right) - e^{\gamma T}\sin\left(\omega(t+T-Q)\right) \right) \times \nonumber  \\
	&& \Theta(t-Q)\nonumber +\frac{I_p}{mT}\frac{1}{\omega^2+\gamma^2}\Theta(t-Q) +  \frac{I_p}{mT} \times  \nonumber \\
	&&  \sum_{k=1}^\infty \frac{2\left(\omega^2+\gamma^2-\omega_k^2\right)\cos\left(\omega_k(t-Q)  \right) + 4 \gamma \omega_k \sin\left(\omega_k(t-Q) \right)}{\left(\omega^2+\gamma^2 -\omega_k^2 \right)^2 + 4\gamma^2 \omega_k^2}\Theta(t-Q). \label{DDCb}
\end{eqnarray}
\end{subequations}

\begin{figure}[tb]
	\includegraphics[scale=0.7]{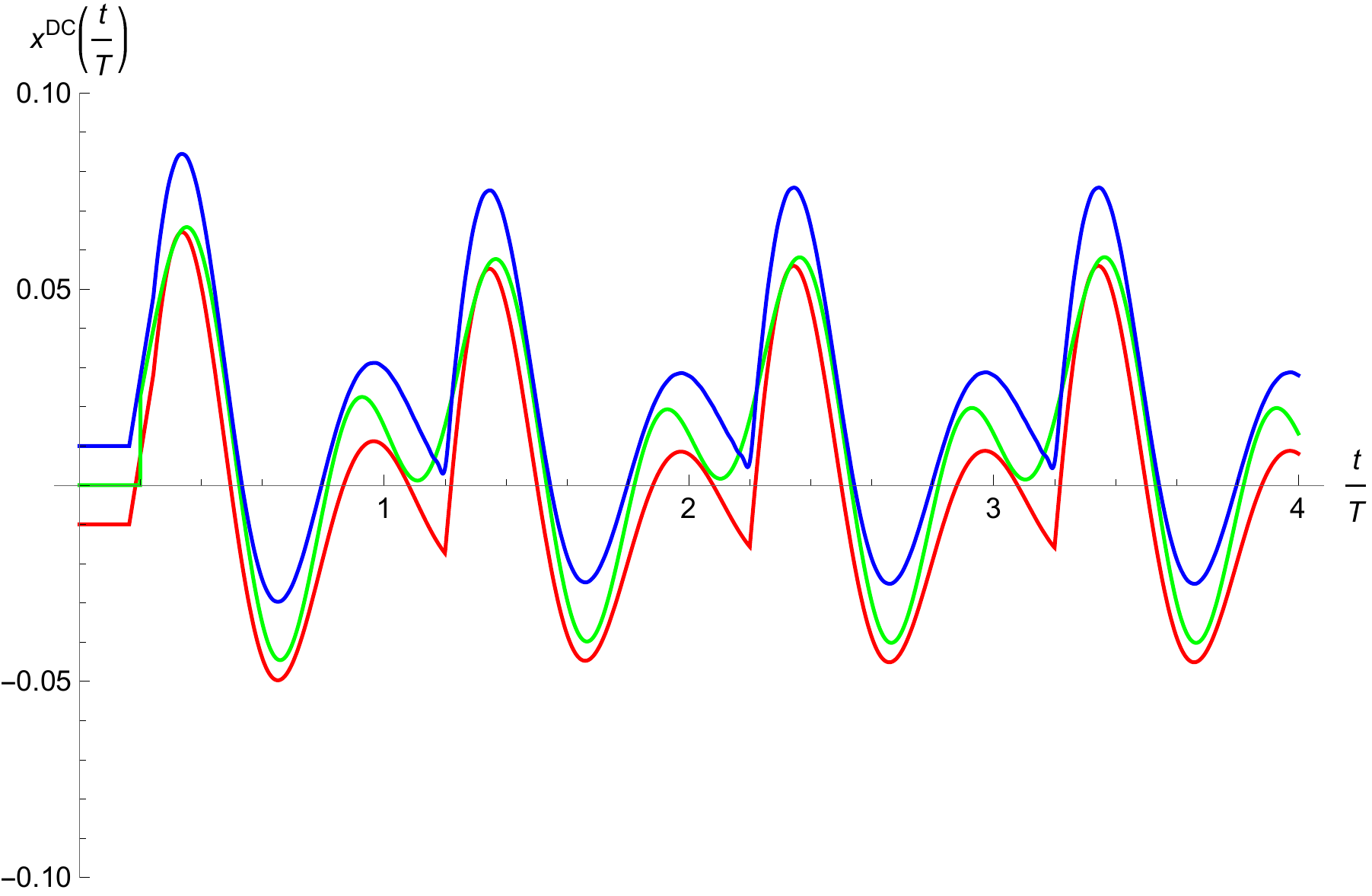}
	\caption{Comparison of particular solutions $x^{\rm DC}(t/T) = x_{\rm TP, HS}^{\rm DC}(t/T)/(I_p/m)$ for periodic solution (red line, Eq.~(\ref{DDCa})) and harmonic solution (green and blue lines, Eq.~(\ref{DDCb})) with the parameters of $\tau/T = 0.001$ and $Q/T= 0.2$, $\gamma = 2$, and $\omega = 10$. The blue and red curves are intentionally shifted by +0.01 and - 0,01, respectively, to see clearly their features. The cut-off orders for green and blue lines are, respectively, $k_c=2$ and $k_c = 30$ (see text). Two solutions (red and blue lines) are perfectly overlayed ater $t>0$, but the green one which includes only up to the second-harmonic order of $\omega_{\rm R}$ capture approximately the eccential features of the red line, i.e., amplitudes, oscillation frequency, and damping rate.}\label{Fig2}
\end{figure}

If the system decay time is shorter than the pulse-to-pulse interval $T$, i.e.,  $\tau_0 = 1/\gamma < T$, the contribution of each pulses are compleletely isolated in the time domain due to the term $e^{-\gamma (t -Q)}$ in Eq.~(\ref{DDC}). Then, the system oscillation period $t_0 = 1/\omega = 1/\sqrt{\omega_0^2 - \gamma^2}$ determines the frequency of oscillation within the system decay time $\tau_0$ as can be seen in Fig.~\ref{Fig2}. It is not clear, however, to see this fast decaying  effect from the harmonic solution in Eq.~(\ref{DDCb}), because it is the solution for $N_d\rightarrow\infty$ and the contribution from the infinute number of simple poles $s = s_k = ik\omega_{\rm R}, k = 0, 1,2 \cdots$ are appeared at the second term for $k = 0$ and the third summation terms for $k =1,2,3, \cdots$. However, the numerical solutions of Eq.~(\ref{DDCa}) and Eq.~(\ref{DDCb}) shown in Fig. \ref{Fig2}  clearly demonstrates the similarity between the two solutions for $t>0$. In Fig.~\ref{Fig2}, the blue and red curves are intentionally shifted by +0.01 and - 0,01, respectively, to see clearly their features, and the cut-off order $k_c$ in the harmonic solutions are, respectively, $k_c = 2$ for green line and $k_c = 30$ for the blue line.

We wish to emphasie that the time-periodic solution in Eq.~(\ref{DDCa}) and the harmonic solution in Eq.~(\ref{DDCb}) for large cut-off order (blue line for $k_c=30$) show exactly the same temporal reaponce of the UHO to the Dirac comb driving. But, from the periodic analytical and numerical solutions, it is hard to see how the UHO responds to the harmonic frequency of the repetition rate $\omega_{\rm R}$ that comes from the periodicity of the driving forces. Clearly, the harmonic solutions (green and blue lines in Fig.~\ref{Fig2}) show the response of the UHO system at the harmonic frequewncy $\omega_k = k \omega_R, k = 0, 1, 2, \cdots,$ of the repetition rate, where $\omega_{\rm R} = 1/T$. 

In Fig. \ref{Fig2}, two solutions (red and blue lines) are perfectly overlayed after $t>0$, but the green one which includes only up to the second-harmonic order of $\omega_{\rm R}$ captures closely the eccential features of the red line, i.e., amplitude, oscillation frequency, and damping rate. In other word, temporal behavior of the UHO driven by the Dirac delta comb could be described at least by the second harmonic response of $\omega_{\rm R}$, i.e., the contributions from the harmonic orders of $k = 0, 1$, and 2. Then, $x_{\rm HS}^{\rm DC}(t)$ in Eq.~(\ref{DDCb}) can be simplified for $t > T+Q$ as the follow
\begin{eqnarray}
		x_{\rm HS}^{\rm DC}(t) &\simeq&  \frac{I_p}{mT} \left( \frac{1}{\omega_0^2} +\sum_{k=1}^2 \frac{2\left(\omega^2+\gamma^2-\omega_k^2\right)\cos\left(\omega_k(t-Q)  \right) + 4 \gamma \omega_k \sin\left(\omega_k(t-Q) \right)}{\left(\omega^2+\gamma^2 -\omega_k^2 \right)^2 + 4\gamma^2 \omega_k^2} \right) \nonumber \\
		&=& \frac{I_p}{mT} \Big(\frac{1}{\omega_0^2}+\frac{2}{\sqrt{4 \gamma^2\omega_{\rm R}^2 + (\omega_0^2 - \omega_{\rm R}^2)}}\sin(\omega_{\rm R}(t-Q)+\phi_1) + \nonumber \\
		&&\frac{2}{\sqrt{16 \gamma^2\omega_{\rm R}^2 + (\omega_0^2 -4 \omega_{\rm R}^2)}}\sin(2\omega_{\rm R}(t-Q)+\phi_2) \Big)\Theta(t-Q), \label{DDCSA}
\end{eqnarray}
where $\phi_1 = \arctan\left(\frac{\omega_0^2-\omega_{\rm R}^2}{2\gamma \omega_{\rm R}}\right)$ and $\phi_2 = \arctan\left(\frac{\omega_0^2-4\omega_{\rm R}^2}{4\gamma \omega_{\rm R}}\right)$. From Eq.~(\ref{DDCSA}), we see that the oscillation frequency of the UHO at the asymtotic limit is $\omega \simeq 2\omega_{\rm R}$ and there are roughly two oscillations between $nT$ and $(n+1)T$, as clearly seen also in Fig. \ref{Fig2} at the given system and driving force parameters. Finally, from the approximate harmonic solution Eq.~(\ref{DDCSA}) and Fig.~\ref{Fig2} (green line), we immediately see that most significant features of the temporal behavior of the UHO driven by the Dirac comb at the asymtotic limit can be captured by the first harmonic and the second harmonic oscillation terms near $\omega = \sqrt{\omega_0^2 -\gamma^2} \simeq  2\omega_{\rm R} = 4\pi/T$, and the relation $\omega \simeq 2 \omega_{\rm R}$ relates the system and driving force parameters through the phase shift $\phi_1$ and $\phi_2$. 

\subsection{Square pulse driving \label{sec:NSP}}

In Appendix~\ref{sec:ATS} and Appendix \ref{sec:AHS}, we found two particular solutions for a train of square pulse driving: the time-periodic solution $x_{\rm TP}^{\rm SP}(t)$ given in Eq.~(\ref{XpsolSP}) with $t_n(t) = t - nT-Q$ \cite{Marion09} and the harmonic solution $x_{\rm HS}^{\rm SP}(t)$ given in Eq.~(\ref{FinalSP}) as follows
\begin{subequations}\label{DSP}
	\begin{eqnarray}
		x_{\rm TP}^{\rm SP}(t) 
		&=&\frac{I_p}{2\tau m}\sum_{n=0}^{N_d} \label{DSPa} \Big(\frac{1}{\gamma^2+\omega^2}\left[\Theta(t_n(t)+\tau)-\Theta(t_n(t)-\tau)\right] \nonumber \\ 
		&&-\frac{e^{-\gamma(t_n(t)+\tau)}}{\omega(\gamma^2+\omega^2)} \left[\omega\cos(\omega(t_n(t)+\tau))+\gamma\sin(\omega(t_n(t)+\tau))\right]\Theta(t_n(t)+\tau)\Large) \nonumber  \\
		&&+\frac{e^{-\gamma(t_n(t)-\tau)}}{\omega(\gamma^2+\omega^2)} \left[\omega\cos(\omega(t_n(t)-\tau))+\gamma\sin(\omega(t_n(t)-\tau))\right]\Theta(t_n(t)-\tau)\Large)\big),  \\ 
		x_{\rm HS}^{\rm SP}(t) &=& 
		 x_{\infty}^{\rm SP}(t-Q+\tau)\Theta(t-Q+\tau) - x_{\infty}^{\rm SP}(t-Q-\tau)\theta(t-Q-\tau),\label{DSPb}  \\  
		 \text{where}  \nonumber \\
		 x_{\rm \infty}^{\rm SP}(t) &=& \frac{I_p}{2\tau m}\frac{-4\gamma +(T+2t)(\gamma^2+\omega^2)}{2T(\gamma^2+\omega^2)^2}  -\frac{I_p}{2\tau m}\frac{e^{-\gamma t}} {\omega \left(\gamma ^2+\omega^2\right)
		 	\left(1 +e^{2 \gamma  T}-2e^{\gamma T}\cos (\omega T)\right)}\times \nonumber \\
		 &&\,\,\,\Big(\omega\cos(\omega t) +\gamma \sin(\omega t)- e^{\gamma T}\left(\omega \cos(\omega (t+T))+\gamma\sin (\omega (t+T))\right)\Big) \nonumber \\ 
		 && +\frac{I_p}{2\tau mT}\sum_{k=1}^\infty \frac{-4 \gamma  \omega _k \cos \left(\omega _k t\right)+2 \left(\gamma^2+\omega^2-\omega_k^2\right) \sin \left(\omega _k t\right)}{\omega_k
		 	((\gamma^2+\omega^2-\omega_k^2)^2 +4\gamma^2\omega_k^2)}.\label{DSPc}
	\end{eqnarray}
\end{subequations}

\begin{figure}[tb]
	\includegraphics[scale=0.6]{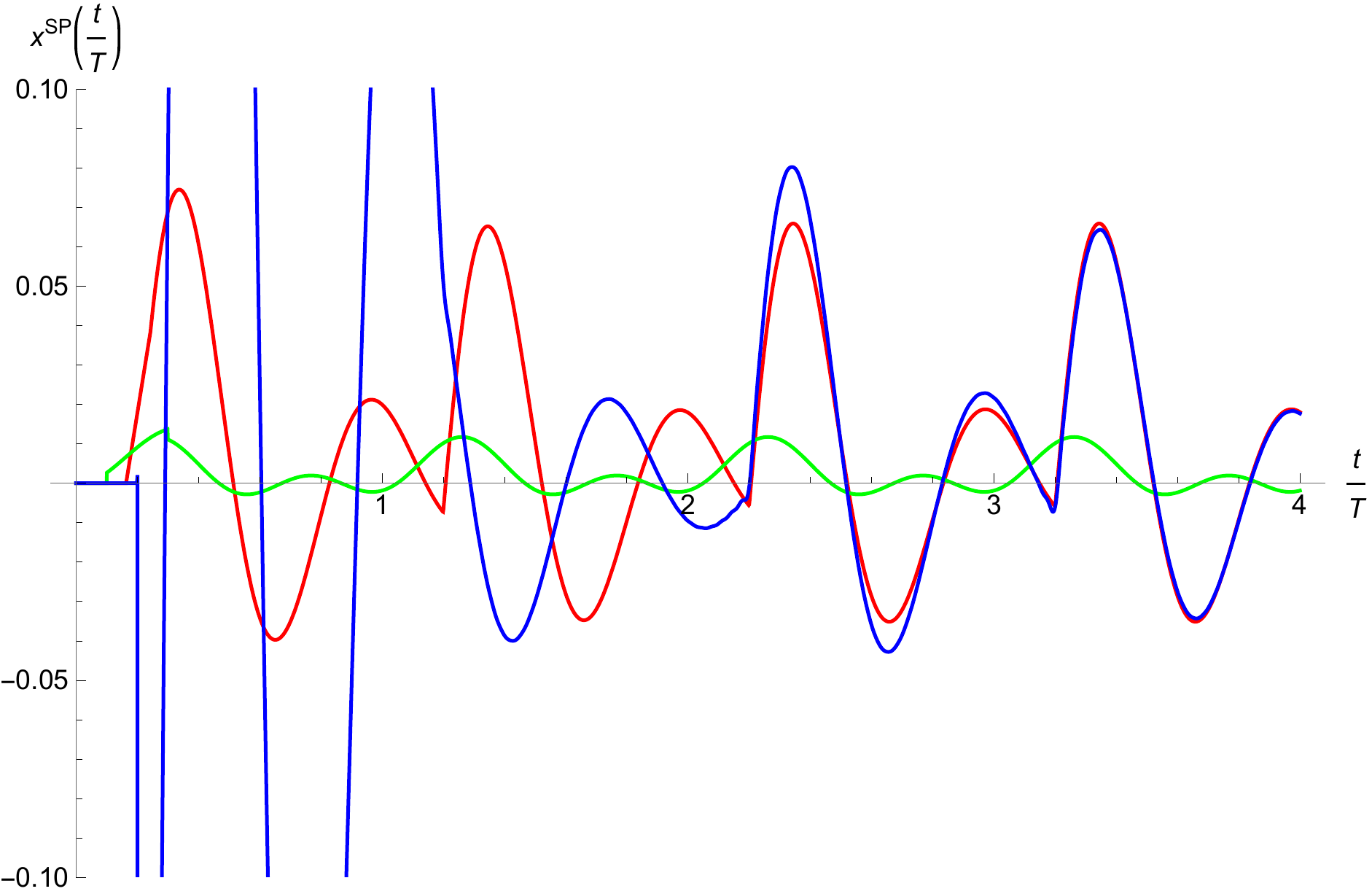}
	\caption{Comparison of particular solutions $x^{\rm SP}(t/T) = x_{\rm TP, HS}^{\rm SP}(t/T)/(I_p/m)$ for periodic solution (red line, Eq.~(\ref{DSPa})) and harmonic solution (green and blue lines, Eq.~(\ref{DSPb})) with the parameters of $\tau/T = 0.001$ and $Q/T= 0.2$, $\gamma = 2$, and $\omega = 10$. The cut-off orders for green and blue lines are, respectively, $k_c=2$ and $k_c = 30$. Two solutions (red and blue lines) are overlayed perfectly only after ater $t>3T$, but the green one which includes up to the second-harmonic order of $\omega_{\rm R}$ does not capture the eccential features of the red line at all for any $t$.}\label{Fig3}
\end{figure}

As in the case of the Dirac comb driving, if the system decay time is shorter than the pulse-to-pulse interval $T$, i.e.,  $\tau_0 = 1/\gamma < T$, the contribution of each pulses are compleletely isolated in the time domain due to the term $e^{-\gamma (t -nT-Q)}$ in Eq.~(\ref{DSPa}). Then, the system oscillation period $t_0 = 1/\omega$ determines the frequency of oscillation within the system decay time $\tau_0$  as shown in Fig.~\ref{Fig3}. Similary to the case of the Dirac comb driving, it is not clear to see this fast decaying effect from the harmonic solution in Eq.~(\ref{DSPb}). We see in Fig.~\ref{Fig3}, however, that the numerical solutions of Eq.~(\ref{DSPa}) and Eq.~(\ref{DSPb}) demonstrate how the UHO system reponds to the train of square pulses differently for the first few pulses. The time-periodic solution clearly show the individual pulse contribution as expected similar to the Dirac comb driving, but the harmonic solution deviates significantly for during first few pulses, but soon approaches the same solution as the time-periodic one after $t>3T$. We attibute this difference due to the feature of square pulses that have impulse responces spectrum at the raising and falling edges, since it needs much higher harmonics of $\omega_{\rm R}$ compared to the Dirac comb pulses. 

It is also clear that the time-periodic solution in Eq.~(\ref{DSPa}) shows the temporal reaponce of the UHO to the trian of square pulse driving, but hard to see how the UHO responds to the harmonic frequency of the repetition rate $\omega_{\rm R}$ that comes from the periodicity of the driving forces. Furthermore, as can be seen Fig. \ref{Fig3}, which is quite different from Dirac comb driving, the harmonic solution with $k_c = 2$ (green line) does not capture the major features of the time-periodic solution. It means that the harmonic solution needs much higher cut-off harmonic order, e.g., $k_c = 30$ (blue line), to make the harmonic solution hehaves close to the time-periodic solution after $t>3T$.

\subsection{Gausian pulse driving \label{sec:NGP}}

In Appendix ~\ref{sec:ATS} and Appendix \ref{sec:AHS}, we found two particular solutions for a train of Gaussian pulse driving \cite{Wright04,Gao13,Gao14,Gao15,Hassan20,Alharbey21}: the time-periodic solution $x_{\rm TP}^{\rm GP}(t)$ given in Eq.~(\ref{XpsolGP}) and the harmonic solution $x_{\rm HS}^{\rm GP}(t)$ given in Eq.~(\ref{FinalGP}) as follows
\begin{subequations}\label{DGP}
	\begin{eqnarray}
		x_{\rm TP}^{\rm GP}(t) 
		&=&\frac{I_p}{m\omega} e^{-\frac{1}{2}\tau^2(\omega^2-\gamma^2)} \sum_{n=0}^{N_d} e^{-\gamma (t-nT-Q)}
		\Big(\sqrt{2}a \tau \omega\cos(\omega(t-nT-Q-\gamma\tau^2)) \nonumber \\ 	\label{DGPa}
		&&-4\sin(\omega(t-nT-Q-\gamma \tau^2))\Big)\Theta(t-nT-Q), \\ 	
		x_{\rm HS}^{\rm GP}(t) &=& 	 x_{\infty}^{\rm GP}(t-Q)\Theta(t-Q), 	\label{DGPb} \\ 
		\text{where} \nonumber \\ 
	 x_{\infty}^{\rm GP}(t) &=&	
		 \frac{I_p}{mT}\frac{1}{\gamma^2+\omega^2}  + \frac{I_p}{m}e^{-\gamma t} e^{\frac{1}{2}\tau^2(\gamma^2-\omega^2)} \, \frac{\sin(\omega(t-\gamma \tau^2)-e^{\gamma T}\sin(\omega(t+T-\gamma \tau^2)))}{\omega(1+e^{2\gamma T}-2e^{\gamma T}\cos(\omega T))} \nonumber \\
		&&+\frac{2I_p}{mT}\sum_{k=1}^\infty e^{-\frac{1}{2} \tau ^2 \omega _k^2} \, \frac{2 \gamma  \omega _k
			\sin \left(\omega _k t\right)+\left(\gamma ^2+\omega^2-\omega _k^2\right) \cos \left( \omega _k t\right)}{
			(\gamma^2+\omega^2-\omega_k^2)^2 +4\gamma^2\omega_k^2}.
		\label{DGPc}
	\end{eqnarray}
\end{subequations}
We note here that the constant $a$ in Eq. (\ref{DGPa}) is found emperically as described in Appendix \ref{sec:AHS}. In short, as discussed in Eq.~(\ref{LTGInfty}), for the UHO with $\tau < T$ and $T,Q > \tau_0 = 1/\gamma$, the complex argument of ${\rm erf}(z)$ at $s = s_{\pm}$ has a large real part ${\rm Re}[z] =\frac{n T + Q + \gamma \tau^2}{\sqrt{2}\tau} \gg 1$ and a small imaginary part ${\rm Im}[z] = \frac{\tau\omega}{\sqrt{2}} \ll 1$. In  this case, we found an approximate formula for ${\rm erf}(z)$ as ${\rm erf}(z)\approx 1 \pm i a \frac{\tau\omega}{\sqrt{2}}$ at $s = s_{\pm}$, here $a \simeq \ln(\pi) \simeq 1.15$ is a constant valid for a wide range of $ {\rm Im}[z]$ below 0.5.

\begin{figure}[tb]
	\includegraphics[scale=0.6]{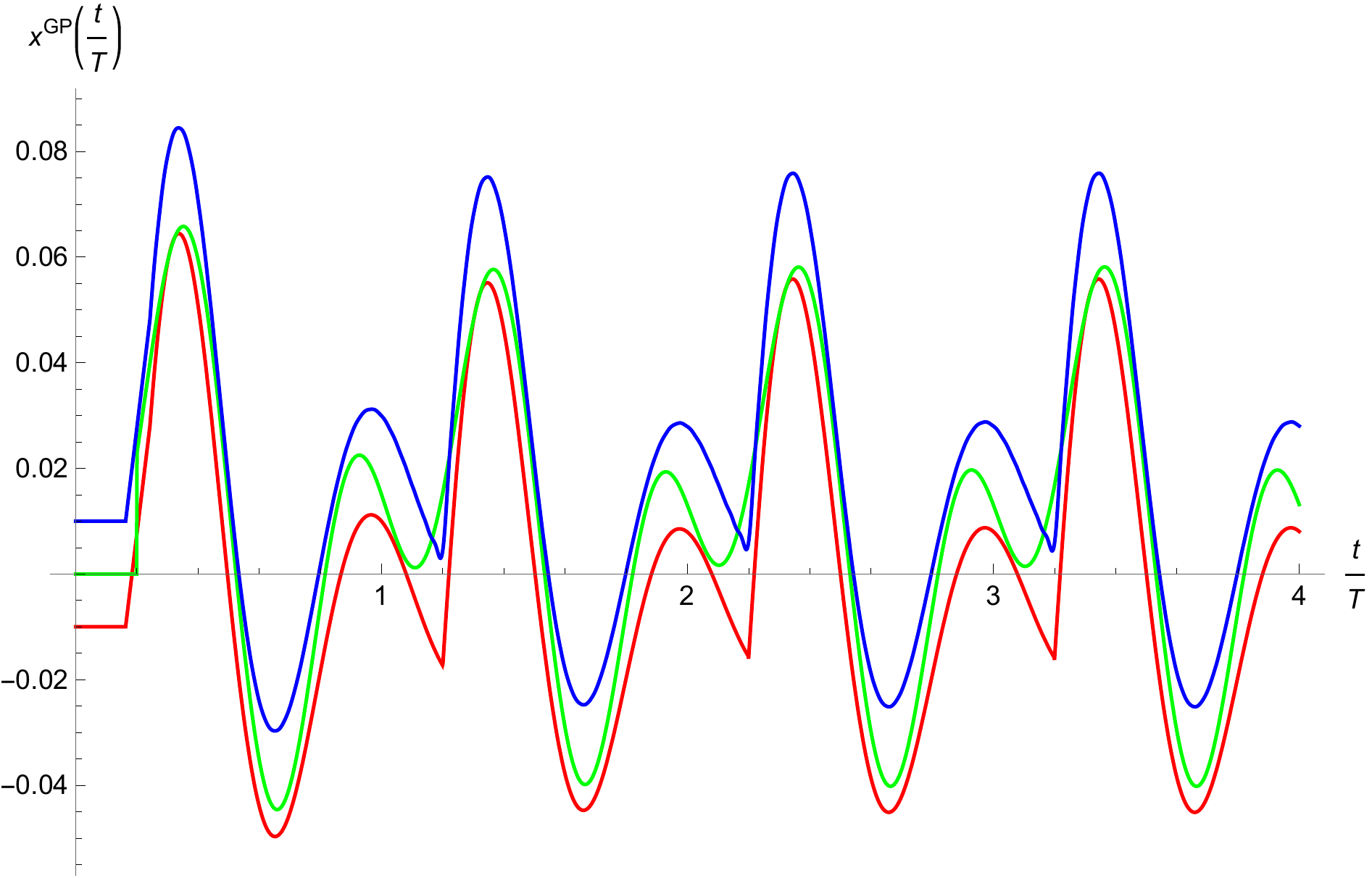}
	\caption{Comparison of particular solutions $x^{\rm GP}(t/T) = x_{\rm TP, HS}^{\rm GP}(t/T)/(I_p/m)$ for periodic solution (red line, Eq.~(\ref{DGPa})) and harmonic solution (green and blue lines, Eq.~(\ref{DGPb})) with the parameters of $\tau/T = 0.001$ and $Q/T= 0.2$, $\gamma = 2$, and $\omega = 10$. The blue and red curves are intentionally shifted by +0.01 and - 0,01, respectively, to see clearly their features. The cut-off orders for green and blue lines are, respectively, $k_c=2$ and $k_c = 30$ (see text). Two solutions (red and blue lines) are perfectly overlayed ater $t>0$, but the green one which includes only up to the second-harmonic order of $\omega_{\rm R}$ capture approximately the eccential features of the red line, i.e., amplitudes, oscillation frequency, and damping rate.}\label{Fig4}
\end{figure}

As in the cases of Dirac comb driving and square pulse driving, if the system decay time is shorter than the pulse-to-pulse interval $T$, i.e.,  $\tau_0 = 1/\gamma < T$, the contribution of each pulses are compleletely isolated in the time domain due to the term $e^{-\gamma (t -nT-Q)}$ in Eq.~(\ref{DGPa}). Then, the system oscillation period $t_0 = 1/\omega$ determines the frequency of oscillation within the system decay time $\tau_0$ as shown in Fig.~\ref{Fig4}. Although, it is not clear to see this fast decaying effect from the harmonic solution in Eq.~(\ref{DGPb}), we can see in Fig.~\ref{Fig4} that the numerical solutions of Eq.~(\ref{DGPa}) and Eq.~(\ref{DGPb}) show how the system reponds to the train of Gaussian pulses. In Fig.~\ref{Fig4}, the blue and red curves are intentionally shifted by +0.01 and - 0,01, respectively, to see clearly their features, and the cut-off order $k_c$ in the harmonic solutions are, respectively, $k_c = 2$ for green line and $k_c = 30$ for the blue line.  From Fig. \ref{Fig4}, we see immediately that  the temporal behavior of the UHO driven by a train of Gaussian pulses can be simplified by the harmonic solutions by taking the harmonic order only up to $k_c = 2$ (green line), this feature is exactly the same as the case of Dirac comb driving shown in Fig. \ref{Fig2}. In addition, the numerical solutions of Eq.~(\ref{DGPa}) and Eq.~(\ref{DGPb}) show clearly the similarity between the two solutions from the very first pulse even for the second harmonic solution (green line). 

In Fig. \ref{Fig4}, two solutions (red and blue lines) are perfectly overlayed after $t>0$, and supringly the green one which includes only up to the second-harmonic order of $\omega_{\rm R}$ already captures all the eccential features of the red line, i.e., amplitude, oscillation frequency, and damping rate. In other words, temporal behavior of the UHO driven by the train of Gaussian pulses would be described by the harmonic solution only up to the second harmonic response of $\omega_{\rm R}$, i.e., the contributions from the harmonic orders of $k = 0, 1$ and $2$. Then, $x_{\rm HS}^{\rm GP}(t)$ in Eq.~(\ref{DGPb}) can be simplified for $t > T+Q$ as the follow
	\begin{eqnarray}\label{DGPHSA}
		x_{\rm HS}^{\rm GP}(t) &=& 	
		\frac{I_p}{mT}\Big(\frac{1}{\omega_0^2}  + 2\sum_{k=1}^{2} e^{-\frac{1}{2} \tau ^2 \omega _k^2}  \times \nonumber \\
		&& \frac{2 \gamma  \omega _k
			\sin \left(\omega _k (t-Q)\right)+\left(\omega_0^2-\omega _k^2\right) \cos \left( \omega _k (t-Q)\right)}{
			(\omega_0^2-\omega_k^2)^2 +4\gamma^2\omega_k^2}\Big)\Theta(t-Q) \nonumber \\
		&=& \frac{I_p}{mT} \Big(\frac{1}{\omega_0^2}+\frac{2 e^{-\frac{1}{2}\tau^2\omega_{\rm R}^2}}{\sqrt{4 \gamma^2\omega_{\rm R}^2 + (\omega_0^2 - \omega_{\rm R}^2)}}\sin(\omega_{\rm R}(t-Q)+\phi_1) + \nonumber \\
		&&\frac{2 e^{-2\tau^2\omega_{\rm R}^2}}{\sqrt{16 \gamma^2\omega_{\rm R}^2 + (\omega_0^2 -4 \omega_{\rm R}^2)}}\sin(2\omega_{\rm R}(t-Q)+\phi_2) \Big)\Theta(t-Q),
	\end{eqnarray}
where $\phi_1 = \arctan\left(\frac{\omega_0^2-\omega_{\rm R}^2}{2\gamma \omega_{\rm R}}\right)$ and $\phi_2 = \arctan\left(\frac{\omega_0^2-4\omega_{\rm R}^2}{4\gamma \omega_{\rm R}}\right)$ as same as for the Dirac comb driving shown in Eq.~(\ref{DDCSA}).  From Fig.~(\ref{Fig4}) and Eq.~(\ref{DGPHSA}), we see that the oscillation frequency of the UHO at the asymtotic limit is $\omega \simeq 2\omega_{\rm R}$, as same as in the case of Dirac comb driving. From Eq.~(\ref{DGPHSA}) and Fig.~\ref{Fig4} (green line), we immediately see that most significant features of the temporal behavior of the UHO driven by the Dirac comb at the asymtotic limit can be captured by the first harmonic and the second harmonic oscillation terms near $\omega = \sqrt{\omega_0^2 -\gamma^2} \simeq 2 \omega_{\rm R} = 4\pi/T$, and the relation $\omega \simeq 2 \omega_{\rm R}$ relates the system and driving force parameters through the phase shift $\phi_1$ and $\phi_2$. Finally, all the system parameters of the UHO, i.e., $\gamma, \omega$, thus the natural frequency $\omega_0$ can now be determined experimentally by nonlinear curve fitting the experimental data to the analytical solutions given in Eqs.~(\ref{DGPa}), (\ref{DGPb}),  and (\ref{DGPHSA}) \cite{Wright04,Gao13,Gao14,Gao15,Hassan20}. 

As already discussed in Appendix \ref{sec:ATS} that the time-periodic solutions of the UHO driven by a Dirac comb and a train of short Gausian pulses are exactly same when $\tau \rightarrow0$. In this section, we also prove that the asymtotic behaviors of the harmonic solutions for the UHO driven by the Dirac comb and trian of short Gaussian pulses are very similar as can be seen in Eqs.~(\ref{DDCSA}) and (\ref{DGPHSA}) and Figs. \ref{Fig2} and \ref{Fig4}. In particular, since $2 \tau^2\omega_{\rm R}^2 \ll 1$ in Eq.~(\ref{DGPHSA}), i.e., $T \gg \tau$, which is the condition for an underdamped oscillation driven by a short Gaussian pulses, two asymtotic harminic solutions in Eqs.~(\ref{DDCSA}) and (\ref{DGPHSA}) becomes identical as expected. Finally, the two independent  analytical solutions, i.e., the time-periodic solution given in Eq. (\ref{DGPa}) and the harmonic solution given in Eqs. (\ref{DGPb}) and (\ref{DGPHSA}), would serve the theoretical and practical models to understand the actual experimental results in various UHO ststems \cite{Dekker81} across various diciples of classical, semiclassical \cite{Yeon87,Kells04,Tayler05,Choi08,Tang08,Marion09,Wright04,Gao13,Zech21}, and quantum harmonic oscillators driven by a real train of short and/or ultra-short periodic Gaussian pulse perturbations \cite{Gzyl83,Um87,Um02,Mudde03,Goldman15,Prado17,Hassan20}.

\section{Summary \label{sec:Summary}}

In symmary, we presented two different analytical solutions: {\it time-periodic solutions} in Appendix \ref{sec:ATS} and {\it harmonic solutions} in Apendix {\ref{sec:AHS}, for the one-dimensional classical UHO driven by three different trains of short pulses, i.e, a Dirac comb, a train of square pulses, and a train of Gaussian pulses with the same pulse width $2\tau$ and pulse-to-pulse time interval $T$. Two solutions for square and Gaussian pulses approach to that of the Dirac comb when $2\tau \rightarrow 0$ as expected. In particular, the harmonic solutions for Dirac comb and Gaussian pulses would be expressed approximately with harmonic terms of the repetition frequency $\omega_{\rm R} = 2\pi/T$ up to the second harmonics. The Dirac comb solutions provide simple theoretical models but hard to apply to simulate the experimental data since it assumes a zero pulse width. The square pulse solutions, however, are more realistic, but still they are not practical since it interacts with the system with flat interaction strength during the pulse width $2\tau$. Finally, the Gaussian pulse solutions are practical analytical solutions those can be directly applicable to  determine experimentally the system parameters by nonlinear curve fitting the experimental data to the presented analytical formulae, e.g., the underdamped oscillation frequency $\omega$, the natural frequency $\omega_0$, and the damping rate $\gamma$, for various parameter regimes of the driving parameters of the pulse width $2\tau$ and pulse repetition rate $\omega_{\rm R}$. 
	
We envision that the presented solutions would expand the current understandings of the periodically driven UHOs in various time scales, for example in a short, intermediate, and asymtotic time scales, in the classical, semiclassical, and quantum science disciplines.


\newpage
\begin{acknowledgments}
This work was supported by NRF-2019R1A2C2009974. T.H.Y. was supported in part by IBS-R023-D1.
\end{acknowledgments}

\appendix

\section{Time-periodic solutions \label{sec:ATS}}

\subsection{Dirac delta comb driving \label{sec:DC}}

In this Appendix \ref{sec:ATS}, we model the Dirac delta comb (DC) \cite{Marion09} as a train of periodic Dirac delta functions with the period $T$ and has a limiting zero pulse width but has the normalized area to be one as same as the square pulses such that 
\begin{eqnarray}\label{Comb}
	F_{\rm d}^{\rm DC}(t,T,Q) &=& \lim_{\tau\rightarrow 0} F_0 2\tau \sum_{n=0}^{N_d}  \frac{1}{2\tau}\left(\Theta(t- nT-Q + \tau) - \Theta(t- nT-Q -\tau)\right) \nonumber \\
	&=& I_p\sum_{n=0}^{N_d} \delta (t - nT - Q),
\end{eqnarray}
where $\Theta(t)$ is the Heaviside unit step function, $\delta(t)$ being the Dirac delta function, $I_p =\int_{-\infty}^\infty F_d(t)dt = F_0 2\tau$, being the impulse delivered to the system by a single pulse, and $F_0$ being the peak force. Now, in order to obtain the particular solution in Eq.~(\ref{ILsolxp}), we need to find $\mathcal{F}_{\rm p}^{\rm DC}(s)$ by taking Laplace transforn of Eq.~(\ref{Comb}) such that
\begin{equation}\label{LFCs}
	\mathcal{F}_{\rm p}^{\rm DC}(s) =I_p \sum_{n=0}^{N_d}
	\int_0^\infty e^{-st}  \delta\left(t - nT - Q\right) dt = I_p\sum_{n=0}^{N_d} e^{-s(nT + Q)}.
\end{equation}

Thus, the particular solution given in Eq.~(\ref{ILsolxp}) for the Dirac delta comb driving can be obtained from the inverse Laplace transform of Eq.~(\ref{LFCs}) as
\begin{equation}\label{xpsol}
	x_{\rm p}^{\rm DC} (t,T,Q) = \frac{I_p}{m}\sum_{n=0}^{N_d}\mathcal{L}^{-1}[\mathcal{H}^{\rm DC}(s)e^{-(nT+Q)s}],
\end{equation}
where $\mathcal{H}^{\rm DC}(s) = \mathcal{H}(s)$ in Eq.~(\ref{TRF}).
To perform the inverse Laplace transform as well as the sum over $n$ in Eq.~(\ref{xpsol}) that includes the exponential term with the first moment in $s$, representing the periodic delta-kick driving, we can take either (1) the inverse Laplace transform $\mathcal{L}^{-1}[\mathcal{H}^{\rm DC}(s)e^{-(nT+Q)s}]$ first for fixed $n$, and then sum the contribution for all $n$ (time-periodic solutions in Appendix \ref{sec:ATS}), or (2) take the geometric sum $\sum_{n=0}^{N_d}e^{-n T s}$ first and then take the inverse Laplace transform of the intermediate result as the final step (harmonic solutions in Appendix \ref{sec:AHS}) \cite{Tayler05}. 

The time-shifting property of the Laplace transform \cite{Afken13} reads for $b>0$
\begin{equation}\label{tshift}
	\mathcal{L}\left[f(t-b)\Theta(t-b)\right]= e^{-bs}\mathcal{F}(s) \,\, \text{or} \,\,
	f(t-b)\Theta(t-b) = \mathcal{L}^{-1} \left[e^{-bs}\mathcal{F}(s)\right],
\end{equation} 
where $\mathcal{L}[f(t)] = \mathcal{F}(s)$. Thus, from Eqs.~({\ref{ILTFGs}) and (\ref{xpsol}) and time-shifting property of the Laplace transform in Eq.~(\ref{tshift}), the particular solution $x_{\rm p}^{\rm DC} (t,T,Q) $ can be simply obtained as
	\begin{eqnarray}\label{XpsolDC}
		x_{\rm p}^{\rm DC} (t,T,Q) &=& \frac{I_p}{m}\sum_{n=0}^{N_d} h(t-nT-Q)\Theta(t-nT-Q) \nonumber \\
		&=&\frac{I_p}{m\omega}\sum_{n=0}^{N_d} e^{-\gamma (t-nT-Q)}\sin(\omega(t-nT-Q))\Theta(t-nT-Q).
	\end{eqnarray}
	
	\subsection{Train of square pulse driving \label{sec:SP}}
	
	A train of square pulses \cite{Marion09} with period $T$, pulse width $2\tau$, and time shift $Q$ is given already in Eq.~(\ref{Comb}) as
	\begin{equation}\label{SP}
		F_{\rm d}^{\rm SP}(t,T,\tau,Q) = I_p\sum_{n=0}^{N_d}\frac{1}{2\tau}\left(\Theta(t - nT - Q + \tau) - \Theta(t - nT - Q - \tau)\right).
	\end{equation}
	
	In order to use the time-shift property in Eq.~(\ref{tshift}), we take Laplace transform of Eq.~(\ref{SP}) such that
	\begin{eqnarray}\label{LSP}
		\mathcal{F}_{\rm d}^{SP}(s) &=& \frac{I_p}{2\tau}\sum_{n=0}^{N_d} \int_0^\infty  e^{-st}  (\Theta(t-nT-Q+\tau)-\Theta(t-nT-Q-\tau)) dt \nonumber \\
		&=& \frac{I_p}{2\tau}\sum_{n=0}^{N_d}\frac{1}{ s}\left(e^{-(nT+Q-\tau)s} - e^{-(nT+Q+\tau)s}\right).
	\end{eqnarray}
	From Eqs.~(\ref{ILsolxp}) and (\ref{LSP}), we can now obtain the particular solution $x_{\rm p}^{\rm SP}(t,T,\tau,Q)$ as the follow
	\begin{eqnarray}\label{SPxpsol}
		x_{\rm p}^{\rm SP}(t,T,\tau,Q) 
		&=& \frac{I_p}{2\tau m}\sum_{n=0}^{N_d} \mathcal{L}^{-1}\left[\frac{1}{s(s-s_+)(s-s_-)}(e^{-(nT+Q-\tau)s} - e^{-(nT+Q+\tau)s})\right] \nonumber \\
		&=& \frac{I_p}{2\tau m}\sum_{n=0}^{N_d} \mathcal{L}^{-1} \left[ \mathcal{H}^{\rm SP}(s)(e^{-(nT+Q-\tau)s} - e^{-(nT+Q+\tau)s})\right],
	\end{eqnarray}
	where 
	\begin{equation}\label{HSPs}
		\mathcal{H}^{\rm SP}(s) = \frac{1}{s(s-s_+)(s-s_-)}.
	\end{equation}
	Now, since  $\mathcal{H}^{\rm SP}(s)$ has three simple poles, $s = 0, s = s_{\pm}$, $h^{\rm SP}(t,T,\tau,Q)$ can be easily obtained as the follow
	\begin{eqnarray}\label{ILGSP}
		h^{\rm SP}(t,T,\tau,Q) &=& \mathcal{L}^{-1}\left[\frac{1}{s(s-s_+)(s-s_-)}\right] \nonumber \\
		&=& \frac{1}{\gamma^2+\omega^2} \left(1-\frac{e^{-\gamma t}}{\omega}\left(\omega\cos(\omega t)+ \gamma \sin(\omega t) \right)\right).
	\end{eqnarray}
	From Eqs.~(\ref{SPxpsol}) and (\ref{ILGSP}) and applying the time-shift property of the Laplace transform in Eq.~(\ref{tshift}), we can finally obtain the particular solution $x_{\rm p}^{\rm SP}(t,T,\tau,Q) $ with the compact notation of $t_n(t) = t-nT-Q$ as the follow
	\begin{eqnarray}\label{XpsolSP}
		&&x_{\rm p}^{\rm SP}(t,T,Q,\tau)  = \frac{I_p}{2\tau m}\sum_{n=0}^{N_d} \Big(\frac{1}{\gamma^2+\omega^2}\left[\Theta(t_n(t)+\tau)-\Theta(t_n(t)-\tau)\right]\nonumber \\ 
		&&-\frac{e^{-\gamma(t_n(t)+\tau)}}{\omega(\gamma^2+\omega^2)} \left[\omega\cos(\omega(t_n(t)+\tau))+\gamma\sin(\omega(t_n(t)+\tau))\right]\Theta(t_n(t)+\tau)\Large) \nonumber \\
		&&+\frac{e^{-\gamma(t_n(t)-\tau)}}{\omega(\gamma^2+\omega^2)} \left[\omega\cos(\omega(t_n(t)-\tau))+\gamma\sin(\omega(t_n(t)-\tau))\right]\Theta(t_n(t)-\tau)\Large)\big).
	\end{eqnarray}
	By taking the limit as $\tau\rightarrow0$ of Eq.~(\ref{XpsolSP}), we immediately see that Eq.~(\ref{XpsolSP}) becomes exactly the same as Eq.~(\ref{XpsolDC}) as it should be.
	
	\subsection{Train of Gaussian pulse driving \label{sec:Gaussian}}
	
	A train of Gaussian pulses \cite{Hassan20} with the pulse-to-pulse period $T$, the full width of $2\sqrt{2}\tau$ (between two $e^{-1}$ points), and the time shift $Q$ can be written as
	\begin{equation}\label{GP}
		F_{\rm d}^{\rm GP}(t,T,\tau,Q) = I_p \sum_{n=0}^{N_d}\frac{1}{\sqrt{2\pi}\tau}e^{- \frac{(t-nT-Q)^2}{2\tau^2}},
	\end{equation}
	where a single Gaussian pulse at $t = Q$ has unit area, i.e., $\int_{-\infty}^\infty \frac{1}{\sqrt{2\pi}\tau}e^{- \frac{(t-Q)^2}{2\tau^2}} dt = 1$ and $I_p = \int_{-\infty}^\infty F_d^{\rm GP}(t)dt$ is the impulse delivered by a single Gaussian pulse. In order to use the time-shift property in Eq.~(\ref{tshift}), we take Laplace transform of Eq.~(\ref{GP}) such that
	\begin{eqnarray}\label{LGP}
		\mathcal{H}_{\rm d}^{\rm GP}(s) &=& I_p\sum_{n=0}^{N_d} \int_0^\infty  e^{-st}  \frac{1}{\sqrt{2\pi}\tau}e^{- \frac{(t-nT-Q)^2}{2\tau^2}} dt \nonumber \\
		&=&  I_p\sum_{n=0}^{N_d}  e^{-(nT+Q)s} \frac{1}{2}e^{\frac{1}{2}\tau^2s^2}\left(1+{\rm erf}\left(\frac{nT+Q-\tau^2s}{\sqrt{2}\tau}\right)\right),
	\end{eqnarray}
	where ${\rm erf}(\cdot)$ is the error function with a copmplex argument $z = \frac{nT + Q - \tau^2 s}{\sqrt{2}\tau}$. 
	
	Now, for the case when $T, Q > \tau_0=1/\gamma$ as well as fast system decay time, i.e., $\tau_0 < 1$ s, the functoin ${\rm erf}(z)$ in Eq.~(\ref{LGP}) may be approximated to be ${\rm erf}(z) \approx 1$ so that Eq.~(\ref{LGP}) becomes
	\begin{equation}\label{LTGInfty}
		\mathcal{H}_{\rm d}^{\rm GP}(s) \simeq   I_p\sum_{n=0}^{N_d}  e^{-(nT+Q)s} e^{\frac{\tau^2s^2}{2}}.
	\end{equation}
	The error committed by this step is rather small, given the fast decay of the Gaussian function for $T > \tau_0$, expecially for a train of short Gaussian pulses. Therefore, those parameters shall be two key control parameters enabling the study of the system decay (damping) time $\tau_0$ relative to the deriving pulse period $T =  2\pi/\omega_{\rm R}$. In particular, for large $n\gg1$, i.e., at the assymtotic (stationary) time when the homogeneous solution decay out completely, the error committed by this approximation is indeed negligible.
	
	From Eqs.~(\ref{ILsolxp}) and (\ref{LGP}), we can now obtain the particular solution $x_{\rm p}^{\rm GP}(t,T,\tau,Q)$ as the follow
	\begin{equation}\label{GPxpsol}
		x_{\rm p}^{\rm GP}(t,T,\tau,Q) 
		= \frac{I_p}{m}\sum_{n=0}^{N_d} \mathcal{L}^{-1} \left[\mathcal{H}^{\rm GP}(s) e^{-(nT+Q)s} \right],
	\end{equation}
	where 
	\begin{equation}\label{hGPt}
		\mathcal{H}^{\rm GP}(s)=\frac{e^{\frac{\tau^2s^2}{2}}}{2(s-s_+)(s-s_-)}\left(1+{\rm erf}(z)\right).
	\end{equation}
	Here, $\mathcal{H}^{\rm GP}(s)$ in Eq.~(\ref{hGPt}) has two simple poles at $s = s_{\pm}$. In order to apply the time shifting property in Eq.~(\ref{tshift}), we need to take an inverse Laplace transform  of Eq.~(\ref{GPxpsol}), $h^{\rm GP}(t,\tau)$, that has a comlex error function ${\rm erf}(z)$ \cite{Afken13} in the numerator. As discussed in Eq.~(\ref{LTGInfty}), in the UHO with $\tau < T$ and $T,Q > \tau_0 = 1/\gamma$, the complex argument of ${\rm erf}(z)$ at $s = s_{\pm}$ has a large real part ${\rm Re}[z] =\frac{n T + Q + \gamma \tau^2}{\sqrt{2}\tau} \gg 1$ and a small imaginary part ${\rm Im}[z] = \frac{\tau\omega}{\sqrt{2}} \ll 1$. In  this case, we found an approximate formula for ${\rm erf}(z)$ as ${\rm erf}(z)\approx 1 \pm i a \frac{\tau\omega}{\sqrt{2}}$ at $s = s_{\pm}$, here $a \simeq 1.15 \simeq \ln(\pi)$ is a constant valid for a wide range of $ {\rm Im}[z]$ below 0.5.
	Then, the inverse Laplace transform $h^{\rm GP}(t,\tau)$ could be obtained as
	\begin{equation}\label{ILGGP}
		h^{\rm GP}(t,\tau) 
		=  -\frac{1}{4\omega}e^{-\gamma t}e^{-\frac{1}{2}\tau^2(\omega^2-\gamma^2)} \left(\sqrt{2}a \tau \omega\cos(\omega(t-\gamma\tau^2))-4\sin(\omega(t-\gamma \tau^2))\right).
	\end{equation}
	One can easily see that when ${\rm Im}[z] \simeq 0$ in Eq.~(\ref{ILGGP}), it becomes exactly the same as Eq.~(\ref{LTGInfty}). 
	From Eqs.~(\ref{GPxpsol}) and (\ref{ILGGP}) and applying the time-shift property of the Laplace transform in Eq.~(\ref{tshift}), we can finally obtain the particular solution $x_{\rm p}^{\rm GP}(t,T,\tau,Q) $ as the follow
	\begin{eqnarray}\label{XpsolGP}
		x_{\rm p}^{\rm GP}(t,T,Q,\tau) &=& \frac{I_p}{m\omega} e^{-\frac{1}{2}\tau^2(\omega^2-\gamma^2)} \sum_{n=0}^{N_d} e^{-\gamma (t-nT-Q)}
		\Big(\sqrt{2}a \tau \omega\cos(\omega(t-nT-Q-\gamma\tau^2)) \nonumber \\
		&&-4\sin(\omega(t-nT-Q-\gamma \tau^2))\Big)\Theta(t-nT-Q).
	\end{eqnarray}
	By taking the limit as $\tau\rightarrow0$ and $b\rightarrow 0$ of Eq.~(\ref{XpsolGP}), we immediately see that Eq.~(\ref{XpsolGP}) becomes exactly the same as Eq.~(\ref{XpsolDC}) as it should be. Thus, we probe as a byproduct that the Dirac delta-function can be represented either by the unit area square pulse in Eq.~(\ref{SP})  and by the unit area Gaussian pulse in Eq.~(\ref{GP}) at the limit of $\tau\rightarrow0$.
	
	
	\section{Harmonic solutions \label{sec:AHS}}
	
	In this Appendix \ref{sec:AHS}, we consider an alternative way of obtaining the particular solutions for three different driving forces discussed in the previous Appendix \ref{sec:ATS}. The mathematical structures of Eqs.~(\ref{XpsolDC}), Eq.~(\ref{XpsolSP}), and Eq.~(\ref{XpsolGP}) are very similar, indeed they have a common structure that depends on the pulse number $n$ as $\mathcal{L}^{-1}[\mathcal{H}^{{\rm K}}(s)\sum_{n=0}^{N_d}e^{-nTs}]$, where ${\rm K}=\{ \rm DC, SP, GP \}$. The former factor $\mathcal{H}^{\rm K}(s)$ represents the response function of the different driving force ${\rm K}$ in the $s$-domain, while the second factor $\sum_{n=0}^{N_d}e^{-nTs}$ represents the periodicity of the pulse train. In the Appendix \ref{sec:ATS}, we used first the time-shift property of the Laplace transform in Eq.~(\ref{tshift}) to get the solution in time domain $h^{\rm K}(t-nT)\Theta(t-nT)$ for each $n$, where $h^{\rm K}(t) = \mathcal{L}^{-1}[\mathcal{H}^{\rm K}(s)]$, resulting in the simple analytical particular solutins in terms of Heaviside unit step function $\Theta$ at each pulse time $t-nT$, and later sum the contributions of all pulses. 
	
	Here, on the other hand, we use the identity of the finite geometric sum first to obtain the particular solutions in Eq.~(\ref{ILsolxp}),
	\begin{equation}\label{FiniteSum}
		\sum_{n=0}^{N_d}e^{-nTs}  = \frac{1-e^{-(Nd+1)sT}}{1-e^{-s T}} \simeq \frac{1}{1-e^{-Ts}},
	\end{equation}
	as contrary to the Appendix \ref{sec:ATS}, where we take the inverse Laplace transform of each driving term first. In the middle of Eq.~(\ref{FiniteSum}), we use the fact that at the simple poles of the denominator, i.e., $s = s_k = i k \omega_{\rm R}$, where $\omega_{\rm R} = 2\pi k /T, k=0,1,2,\cdots$, the factor $e^{-(Nd+1)s_kT}$ in the numerator becomes zero for relatively large $N_d$. In other words, the system exhibits the asymtotic response after interaction with $N_d \gg 1$ pulses, which is the same as the time scale when the system interaced with an infinite number of pulses. In this way, we can obtain the closed form of analytical expressions of the particular solutions for the case when the system exhibits an asymtotic temporal response after interactions with large number of driving pulses.
	
	The particular solutions to be obtained, however, have infinite sum of harmonic frequencies, $\omega_k$, as will be seen below, due to the infinite number of simple poles of the denominator of Eq.~(\ref{FiniteSum}) at the imaginary axis, i.e., $s = i\omega_k = i k \omega_{\rm R}$. Therefore, to investigate the temporal dependence of the particular solutions numerically, one need to truncate the sum of the harmonic series of $\omega_{\rm R}$ at the suffuciently higher order at which the remaining contributions would be negligible.
	
	\subsection{Dirac delta comb driving \label{sec:ADC}}
	
	From Eq.~(\ref{FiniteSum}) and the Bromwich integral \cite{Afken13}, Eq.~(\ref{xpsol}) can be written from the time-shifting property of the Laplace transform as the follow
	\begin{equation}\label{AxpsolDC}
		x_{\rm p}^{\rm DC}(t,T,Q)
		= \frac{I_p}{m}\mathcal{L}^{-1} \left[ \mathcal{H}_\infty^{\rm DC}(s)e^{-Qs}\right]
		= \frac{I_p}{m} x_\infty^{\rm DC}(t-Q)\Theta(t-Q),
	\end{equation}
	where 
	\begin{equation}\label{xDCIt}
		\mathcal{H}_\infty^{\rm DC}(s)= \frac{1}{(s-s_+)(s-s_-)(1-e^{-Ts})},
	\end{equation} and 
	\begin{equation}\label{GsxDC}
		x_\infty^{\rm DC}(t) = \mathcal{L}^{-1}[\mathcal{H}_\infty^{\rm DC}(s)] 
		= \frac{1}{2\pi i}\int_{\ell-i\infty}^{\ell+i\infty} \frac{e^{st}}{(s-s_+)(s-s_-)(1-e^{-Ts})}ds.
	\end{equation}
	
	In order to calculate the contour integral of Eq.~(\ref{GsxDC}), one can use the Cauthy's complex integral and residue theorems \cite{Afken13}. Since Eq.~(\ref{GsxDC}) has two simple poles at $s=s_{\pm}$ and an infinite number of poles at $s = s_k = i k \omega_{\rm R}, k = 0, \pm1, \cdots$, i.e., at the imaginary axis, we may calculate the residuals separately such as 
	\begin{equation}\label{SolxpSP}
		x_{\infty}^{\rm DC}(t,T,Q) = {\rm Res}[\mathcal{L}^{-1}[\mathcal{H}_{\infty}^{\rm DC}(s)]_{s_+}] + {\rm Res}[\mathcal{L}^{-1}[\mathcal{H}_{\infty}^{\rm DC}(s)]_{s_-}] + \sum_{k = -\infty}^\infty {\rm Res}[\mathcal{L}^{-1}[\mathcal{H}_{\infty}^{\rm DC}(s)]_{s_k}],
	\end{equation}
	where ${\rm Res}[\mathcal{L}^{-1}[\mathcal{H}_{\infty}^{\rm DC}(s)]_{s_k}]$ stands for the residue at $s = s_k$. Since the poles in Eq.~(\ref{GsxDC}) are simple poles, we can use the residue theorem straightlforwardly to obtain  ${\rm Res}[\mathcal{L}^{-1}[\mathcal{H}_{\infty}^{\rm DC}(s)]_{s_+}]$ and  ${\rm Res}[\mathcal{L}^{-1}[\mathcal{H}_{\infty}^{\rm DC}(s)]_{s_-}]$, respectively, as follows
	\begin{subequations}{\label{XsolpCpm}}
		\begin{eqnarray}
			{\rm Res}[\mathcal{L}^{-1}[\mathcal{H}_{\infty}^{\rm DC}(s)]_{s_+}] &=& -\frac{i}{2\omega } \frac{e^{-t(\gamma- i\omega)}}{(1-e^{T(\gamma - i\omega)})},  \\
			{\rm Res}[\mathcal{L}^{-1}[\mathcal{H}_{\infty}^{\rm DC}(s)]_{s_-}]   &=&   \frac{i}{2\omega }\frac{e^{-t(\gamma+ i\omega)}}{(1-e^{T(\gamma + i\omega)})}.
		\end{eqnarray}
	\end{subequations}
	Equation~(\ref{XsolpCpm}) can be added and further simplified to be ${\rm Res}[\mathcal{L}^{-1}[\mathcal{L}_{\rm p}^{\rm C}(s)]_{s_{\pm}}]$ as
	\begin{equation}\label{DCxolpm}
		{\rm Res}[\mathcal{L}^{-1}[\mathcal{L}_{\rm p}^{\rm C}(s)]_{s_{\pm}}] = \frac{e^{-\gamma t}}{\omega(1-2e^{\gamma T}\cos(\omega T)+e^{2\gamma T})}\left(\sin(\omega t) - e^{\gamma T}\sin(\omega(t+T)) \right).
	\end{equation}
	
	Now, the infinite number of simple poles  can be written in three separated terms corresponding for $k = 0, k>0$, and $k <0$, repectively, as follows
	\begin{equation} 
		\sum_{k = -\infty}^\infty {\rm Res}[\mathcal{L}^{-1}[\mathcal{H}_{\infty}^{\rm DC}(s)]_{s_k}] = {\rm Res}[\mathcal{L}^{-1}[\mathcal{H}_{\infty}^{\rm DC}(s)]_{s_0}] + \sum_{k = \pm 1}^{\pm\infty} {\rm Res}[\mathcal{L}^{-1}[\mathcal{H}_{\infty}^{\rm DC}(s)]_{s_k}],
	\end{equation}
	where ${\rm Res}[\mathcal{L}^{-1}[\mathcal{H}_{\infty}^{\rm DC}(s)]_{s_0}], \sum_{k = 1}^\infty {\rm Res}[\mathcal{L}^{-1}[\mathcal{H}_{\infty}^{\rm DC}(s)]_{s_k}]$, and $ \sum_{k = -1}^{-\infty} {\rm Res}[\mathcal{L}^{-1}[\mathcal{H}_{\infty}^{\rm DC}(s)]_{s_k}]$ can be obtained from the Bromwich integral and the residue theorems as follows
	\begin{subequations}{\label{XsolDCkpm}}
		\begin{eqnarray}
			{\rm Res}[\mathcal{L}^{-1}[\mathcal{H}_{\infty}^{\rm DC}(s)]_{s_0}] &=&  \frac{1}{T}\frac{e^{s_0 t}}{(s_0-s_+)(s_0-s_-)} = \frac{1}{T}\frac{1}{\omega^2+\gamma^2},  \label{DCpmm0}\\
			\frac{1}{T}\sum_{k = 1}^\infty {\rm Res}[\mathcal{L}^{-1}[\mathcal{H}_{\infty}^{\rm DC}(s)]_{s_k}]   &=&  \sum_{k=1}^\infty \frac{1}{T}\frac{e^{s_k t}}{(s_k-s_+)(s_k-s_-)}= \sum_{k=1}^\infty h_{kp}(t), \label{DCmmp}\\  
			\sum_{k = -1}^{-\infty} {\rm Res}[\mathcal{L}^{-1}[\mathcal{H}_{\infty}^{\rm DC}(s)]_{s_k}] &=&  \sum_{k=-1}^{-\infty}  \frac{1}{T}\frac{e^{s_k t}}{(s_k-s_+)(s_k-s_-)} =  \sum_{k=1}^\infty h_{km}(t), \label{DCmmm}
		\end{eqnarray}
	\end{subequations}
	where we simplified the limits in Eq.~(\ref{XsolDCkpm}) by using the  L'H\^{o}pital's rule \cite{Afken13} as
	$\lim_{s\rightarrow s_k}\left[\frac{s-s_k}{1-e^{-sT}}\right] = \lim_{s\rightarrow s_k} \frac{1}{Te^{-sT}} = \frac{1}{T}$,
	with $s_k = i\omega_k=ik\omega_{\rm R}$. As a result,  $h_{kp}(t)$ and $h_{km}(t)$ in Eq.~(\ref{DCmmp}) and Eq.~(\ref{DCmmm}) can easily be otained, respectively, as follows
	\begin{subequations}{\label{DCpmm11}}
		\begin{eqnarray}
			h_{kp}(t) &=&  \frac{1}{T}\frac{e^{i  \omega_k t}}{-\omega_k^2+2\gamma i \omega_k + \omega^2+\gamma^2}, \label{DCpmmp}\\
			h_{km}(t) &=& \frac{1}{T}\frac{e^{-i  \omega_k t}}{-\omega_k^2-2\gamma i \omega_k + \omega^2+\gamma^2}. \label{DCpmmm}
		\end{eqnarray}
	\end{subequations}
	Thus, the two terms of Eqs.~(\ref{DCpmmp}) and (\ref{DCpmmm}) can be summed in a compact form as
	\begin{eqnarray}\label{DCsum}
		\sum_{k=-\infty, \,k\ne0}^\infty {\rm Res}[\mathcal{L}^{-1}[\mathcal{H}_{\infty}^{\rm DC}(s)]_{s_k}] &=& \frac{1}{T} \sum_{k=1}^\infty \frac{2\left(\omega^2+\gamma^2-\omega_k^2\right)\cos\left(\omega_kt  \right) + 4 \gamma \omega_k \sin\left(\omega_kt \right)}{\left( \omega_0^2+\gamma^2 -\omega_k^2 \right)^2 + 4\gamma^2 \omega_k^2}.
	\end{eqnarray}
	Finally, we obtained the closed form of the particular solution $x_{\infty}^{\rm DC}(t)$ from Eqs.~(\ref{DCxolpm}), (\ref{DCpmm0}) and (\ref{DCsum}) after applying time shift $t\rightarrow t-Q$ from Eq.~(\ref{AxpsolDC}) as the follow
	\begin{eqnarray}\label{xpt}
		x_{\rm p}^{\rm DC}(t) &=& \frac{I_pe^{-\gamma (t-Q)}}{m\omega(1-2e^{\gamma T}\cos(\omega T)+e^{2\gamma T})}\left(\sin\left(\omega(t-Q)\right) - e^{\gamma T}\sin\left(\omega(t+T-Q)\right) \right) \times \nonumber  \\
		&& \Theta(t-Q)\nonumber +\frac{I_p}{mT}\frac{1}{\omega^2+\gamma^2}\Theta(t-Q) +   \nonumber \\
		&&  \frac{I_p}{mT} \sum_{k=1}^\infty \frac{2\left(\omega^2+\gamma^2-\omega_k^2\right)\cos\left(\omega_k(t-Q)  \right) + 4 \gamma \omega_k \sin\left(\omega_k(t-Q) \right)}{\left(\omega^2+\gamma^2 -\omega_k^2 \right)^2 + 4\gamma^2 \omega_k^2}\Theta(t-Q).
	\end{eqnarray}

	\subsection{Train of square pulse driving \label{sec:ASP}}
	
	From Eq.~(\ref{FiniteSum}) and the Bromwich integral, Eq.~(\ref{SPxpsol})  can be written as 
	\begin{eqnarray}\label{AxpsolSP}
		x_{\rm p}^{\rm SP}(t,T,\tau,Q) &=& \frac{I_p}{2\tau m}\mathcal{L}^{-1}\left[\mathcal{H}^{\rm SP}(s)\frac{1}{1-e^{-Ts}}\left(e^{-(Q-\tau)s}-e^{-(Q+\tau)s}\right)\right] \nonumber \\
		&=& \frac{I_p}{2\tau m}\mathcal{L}^{-1}\left[\mathcal{H}_\infty^{\rm SP}(s)\left(e^{-(Q-\tau)s}-e^{-(Q+\tau)s}\right)\right]\nonumber \\
		&=& x_{\infty}^{\rm SP}(t-Q+\tau)\Theta(t-Q+\tau) - x_{\infty}^{\rm SP}(t-Q-\tau)\Theta(t-Q-\tau),
	\end{eqnarray}
	where 
	\begin{subequations}\label{GSPs}
		\begin{eqnarray}\label{GSPsa}
			\mathcal{H}_\infty^{\rm SP}(s) &=& \frac{1}{s(s-s_+)(s-s_-)(1-e^{-Ts})}\, \text{and}  \, \\
			x_{\infty}^{\rm SP}(t)  &=&   \frac{I_p}{2\tau m}\mathcal{L}^{-1}\left[\mathcal{H}_\infty^{\rm SP}(s)\right].
			\label{GSPsb}
		\end{eqnarray}
	\end{subequations}
	In order to calculate the contour integral of Eq.~(\ref{GSPsb}), one can use the Cauthy's complex integral and residue theorems as same as the case of Dirac comb. Since Eq.~(\ref{GSPsa}) has one second-order pole at $s = 0$, two simple poles at $s= s_{\pm}$, and an infinite number of simple poles at $s = s_k = i \omega_k = i k \omega_{\rm R}, k = \pm1, \pm2, \cdots,$ i.e., at the imaginary axis, we may calculate the residuals separately such that 
	\begin{eqnarray}\label{SolxpSPR}
		x_{\infty}^{\rm SP}(t,T,\tau,Q) &=&  {\rm Res}[\mathcal{L}^{-1}[\mathcal{H}_\infty^{\rm SP}(s)]_{s_0}]+ {\rm Res}[\mathcal{L}^{-1}[\mathcal{H}_\infty^{\rm SP}(s)]_{s_+}] + {\rm Res}[\mathcal{L}^{-1}[\mathcal{H}_\infty^{\rm SP}(s)]_{s_-}] \nonumber \\
		&& + \sum_{k = \pm1}^{\pm\infty} {\rm Res}[\mathcal{L}^{-1}[\mathcal{H}_\infty^{\rm SP}(s)]_{s_k}],
	\end{eqnarray}
	where ${\rm Res}[\mathcal{L}^{-1}[\mathcal{H}_\infty^{\rm SP}(s)]_{s_k}]$ stands for the residue at $s = s_k$. 
	
	Firstly, $ {\rm Res}[\mathcal{L}^{-1}[\mathcal{H}_\infty^{\rm SP}(s)]_{s_0}]$ in Eq.~(\ref{SolxpSPR}) can be calculated from the residue theorem for second-order poles arising from the factor $1/(s(1-e^{-Ts})) $ in Eq.~(\ref{GSPsa}) as the follow.
	\begin{eqnarray}\label{ResSPs0}
		{\rm Res}[\mathcal{L}^{-1}\left[\mathcal{H}_\infty^{\rm SP}(s)\right]_{s_0}] &=&  \lim_{s\rightarrow0}\frac{d}{ds}\left[\frac{e^{st}}{(s-s_+)(s-s_-)}\frac{s}{(1-e^{-Ts})}\right]\nonumber \\
		&=&\frac{1}{s_+s_-}\frac{1}{T}\left(t+\frac{1}{s_+}+\frac{1}{s_-}+\lim_{s\rightarrow0}\frac{1-(1+Ts)e^{-Ts}}{s(1-e^{-Ts})}\right)\nonumber \\
		&=& \frac{-4\gamma +(T+2t)(\gamma^2+\omega^2)}{2T(\gamma^2+\omega^2)^2},
	\end{eqnarray}
	here we used the  L'H\^{o}pital's rule to calculate $\lim_{s\rightarrow0}\frac{s}{1-e^{-Ts}}=\frac{1}{T}$ and $\lim_{s\rightarrow0}\frac{1-(1+Ts)e^{-Ts}}{s(1-e^{-Ts})} = -\frac{T}{2}$.
	
	Secondly, since the poles $s_{\pm}$ in Eq.~(\ref{GSPs}) are simple poles, we can  use the residue theorem straightlforwardly to obtain  ${\rm Res}[\mathcal{L}^{-1}\left[\mathcal{H}_\infty^{\rm SP}(s)\right]_{s_+}]$ and   $ {\rm Res}[\mathcal{L}^{-1}\left[\mathcal{H}_\infty^{\rm SP}(s)\right]_{s_-}]$, respectively, as the follow
	\begin{subequations}{\label{SolSPpm}}
		\begin{eqnarray}
			{\rm Res}[\mathcal{L}^{-1}\left[\mathcal{H}_\infty^{\rm SP}(s)\right]_{s_+}] 
			&=&\frac{i}{2\omega} \frac{ e^{-t (\gamma -i \omega )}}{(\gamma -i \omega) \left(1-e^{T (\gamma - i \omega )}\right)},  \\
			{\rm Res}[\mathcal{L}^{-1}\left[\mathcal{H}_\infty^{\rm SP}(s)\right]_{s_-}]   
			&=&-\frac{i}{2\omega} \frac{ e^{-t (\gamma +i \omega )} }{(\gamma +i \omega
				) \left(1-e^{T (\gamma +i \omega )}\right)}.
		\end{eqnarray}
	\end{subequations}
	Equation~(\ref{SolSPpm}) can be added and further simplified to be ${\rm Res}[\mathcal{L}^{-1}\mathcal{H}_\infty^{\rm SP}(s)]_{s_{\pm}}]$ as
	\begin{eqnarray}\label{SPxolp2}
		&&{\rm Res}[\mathcal{L}^{-1}[\mathcal{H}_\infty^{\rm SP}(s)]_{s_{\pm}}] = -\frac{e^{-\gamma t}} {\omega \left(\gamma ^2+\omega^2\right)
			\left(1 +e^{2 \gamma  T}-2e^{\gamma T}\cos (\omega T)\right)}\times \nonumber \\
		&&\Big[\omega\cos(\omega t) +\gamma \sin(\omega t)- e^{\gamma T}\left[\omega \cos(\omega (t+T))+\gamma\sin (\omega (t+T))\right]\Big].
	\end{eqnarray}

	Now, the infinite number of simple poles those appear only in the imaginary axis of the last term of Eq.~(\ref{SolxpSPR}), i.e., $s_k = i\omega_k= i k \omega_{\rm R}, k=\pm1, \pm2, \cdots$, can be written in two separated terms corresponding for $k>0$, and $k <0$, repectively as the follow
	\begin{equation} 	
		\sum_{k = -\infty,\,k\ne0}^\infty  {\rm Res}[\mathcal{L}^{-1}[\mathcal{H}_\infty^{\rm SP}(s)]_{s_k}] = \sum_{k = 1}^{\infty}  {\rm Res}[\mathcal{L}^{-1}[\mathcal{H}_\infty^{\rm SP}(s)]_{s_k}]+\sum_{k = -1}^{-\infty}  {\rm Res}[\mathcal{L}^{-1}\left[\mathcal{H}_\infty^{\rm SP}(s)\right]_{s_k}],
	\end{equation}
	where $  \sum_{k = 1}^\infty  {\rm Res}[\mathcal{L}^{-1}[\mathcal{H}_\infty^{\rm SP}(s)]_{s_k}]$ and $ \sum_{k = -1}^{-\infty} {\rm Res}[\mathcal{L}^{-1}[\mathcal{H}_\infty^{\rm SP}(s)]_{s_k}]$ can be obtained from the Bromwich integral and the residue theorems as follows
	\begin{subequations}{\label{XsolpSP}}
		\begin{eqnarray}
			\sum_{k = 1}^\infty  {\rm Res}[\mathcal{L}^{-1}[\mathcal{H}_\infty^{\rm SP}(s)]_{s_k}]   
			&=& \frac{1}{T}\sum_{k=1}^\infty \frac{e^{i\omega_k t}}{i\omega_k(-\omega_k^2+2i\gamma\omega_k + \gamma^2+\omega^2)}, \label{SPmmpp}\\ 
			\sum_{k = -1}^{-\infty}  {\rm Res}[\mathcal{L}^{-1}[\mathcal{H}_\infty^{\rm SP}(s)]_{s_k}]
			&=& \frac{1}{T}\sum_{k=1}^\infty \frac{e^{-i\omega_k t}}{-i\omega_k(-\omega_k^2-2i\gamma \omega_k + \gamma^2+\omega^2)}. \label{SPmmmm}
		\end{eqnarray}
	\end{subequations}
	Thus, the two terms of Eqs.~(\ref{SPmmpp}) and (\ref{SPmmmm}) can be summed in a compact form as the follow
	\begin{equation}\label{SPsum2}
		\sum_{k=-\infty, \,k\ne0}^\infty {\rm Res}[\mathcal{L}^{-1}[\mathcal{H}_\infty^{\rm SP}(s)]_{s_k}] =  \frac{1}{T}\sum_{k=1}^\infty \frac{-4 \gamma  \omega _k \cos \left(\omega _k t\right)+2 \left(\gamma^2+\omega^2-\omega_k^2\right) \sin \left(\omega _k t\right)}{\omega_k
			((\gamma^2+\omega^2-\omega_k^2)^2 +4\gamma^2\omega_k^2)}.
	\end{equation}
	Therefore, we obtained the final closed form of the particular solution $x_{\rm p}^{\rm SP}(t)$ from Eqs.~(\ref{ResSPs0}), (\ref{SPxolp2}), and (\ref{SPsum2}) as
	\begin{eqnarray}\label{SPpsolSP}
		x_{\rm \infty}^{\rm SP}(t) &=& \frac{I_p}{2\tau m}\frac{-4\gamma +(T+2t)(\gamma^2+\omega^2)}{2T(\gamma^2+\omega^2)^2} -\frac{I_p}{2\tau m}\frac{e^{-\gamma t}} {\omega \left(\gamma ^2+\omega^2\right)
			\left(1 +e^{2 \gamma  T}-2e^{\gamma T}\cos (\omega T)\right)}\times \nonumber \\
		&&\,\,\,\Big(\omega\cos(\omega t) +\gamma \sin(\omega t)- e^{\gamma T}\left(\omega \cos(\omega (t+T))+\gamma\sin (\omega (t+T))\right)\Big) \nonumber \\
		&& +\frac{I_p}{2\tau mT}\sum_{k=1}^\infty \frac{-4 \gamma  \omega _k \cos \left(\omega _k t\right)+2 \left(\gamma^2+\omega^2-\omega_k^2\right) \sin \left(\omega _k t\right)}{\omega_k
			((\gamma^2+\omega^2-\omega_k^2)^2 +4\gamma^2\omega_k^2)}.
	\end{eqnarray}
	Finally, from Eq.~(\ref{AxpsolSP}), the analytical expression of the particular solution can be written as the follow
	\begin{equation}\label{FinalSP}
		x_{\rm p}^{\rm SP}(t,T,\tau,Q) = x_{\infty}^{\rm SP}(t-Q+\tau)\Theta(t-Q+\tau) - x_{\infty}^{\rm SP}(t-Q-\tau)\theta(t-Q-\tau),
	\end{equation}
	where $x_{\infty}^{\rm SP}(t)$ is given in Eq.~(\ref{SPpsolSP}). By taking the limit as $\tau\rightarrow0$ of Eq.~(\ref{FinalSP}), we immediately see that Eq.~(\ref{FinalSP}) becomes exactly the same as Eq.~(\ref{xpt}) as it should be.

	\subsection{Train of Gaussian pulse driving \label{sec:AGP}}
	
	From Eq.~(\ref{FiniteSum}) and the Bromwich integral, Eq.~(\ref{GPxpsol})  can be written as 
	\begin{eqnarray}\label{AxpsolGP}
		x_{\rm p}^{\rm GP}(t,T,\tau,Q) &=& \frac{I_p}{m}\mathcal{L}^{-1}\left[\mathcal{H}^{\rm GP}(s)\frac{1}{1-e^{-Ts}}e^{-Qs}\right] \nonumber \\
		&=& \frac{I_p}{m}\mathcal{L}^{-1}\left[\mathcal{H}_\infty^{\rm GP}(s)e^{-Qs}\right]\nonumber \\
		&=& x_{\infty}^{\rm GP}(t-Q)\Theta(t-Q),
	\end{eqnarray}
	where 
	\begin{subequations}\label{GGPs}
		\begin{eqnarray}\label{GGPsa}
			\mathcal{H}_\infty^{\rm GP}(s) &=&  \frac{ e^{\frac{\tau^2s^2}{2}}}{(s-s_+)(s-s_-)(1-e^{-Ts})}  \,\text{and}\, \\
			x_{\infty}^{\rm GP}(t)  &=& \frac{I_p}{m} \mathcal{L}^{-1}\left[\mathcal{H}_\infty^{\rm GP}(s)\right].
			\label{GGPsb}
		\end{eqnarray}
	\end{subequations}
	In order to calculate the contour integral of Eq.~(\ref{GGPsb}), one can use the Cauthy's complex integral and residue theorems as same as the case of Dirac comb and square pulses. Since Eq.~(\ref{GGPsa}) has one simple pole at $s=0$, two simple poles at $s= s_{\pm} $, and an infinite number of simple poles at $s = s_k = i k\omega_{\rm R}, k =  \pm1, \pm2, \cdots,$ i.e., at the imaginary axis, we may calculate the residuals separately such that 
	\begin{eqnarray}\label{SolxpGPR}
		x_{\rm p}^{\rm GP}(t,T,\tau,Q) &=& {\rm Res}[\mathcal{L}^{-1}[\mathcal{H}_\infty^{\rm GP}(s)]_{s_0}] +  {\rm Res}[\mathcal{L}^{-1}[\mathcal{H}_\infty^{\rm GP}(s)]_{s_+}] + {\rm Res}[\mathcal{L}^{-1}[\mathcal{H}_\infty^{\rm GP}(s)]_{s_-}] \nonumber \\
		&& + \sum_{k = \pm1}^{\pm\infty} {\rm Res}[\mathcal{L}^{-1}[\mathcal{H}_\infty^{\rm GP}(s)]_{s_k}],
	\end{eqnarray}
	where ${\rm Res}[\mathcal{L}^{-1}[\mathcal{H}_\infty^{\rm GP}(s)]_{s_k}]$ stands for the residue at $s = s_k$ for $k \in \mathbb{Z}$. 
	
	Firstly, $ {\rm Res}[\mathcal{L}^{-1}[\mathcal{H}_\infty^{\rm SP}(s)]_{s_0}]$ in Eq.~(\ref{SolxpGPR}) can be calculated from the residue theorem for simple pole from the factor $1/(1-e^{-Ts}) $ in Eq.~(\ref{GGPsa}) as the follow
	\begin{equation}
		{\rm Res}[\mathcal{L}^{-1}[\mathcal{H}_\infty^{\rm GP}(s)]_{s_0}]  = \frac{1}{T} \frac{1}{\gamma^2+\omega^2}. \label{GPRes0}
	\end{equation}
	
	Secondly, since the poles $s_{\pm}$ in Eq.~(\ref{GGPs}) are simple poles, we can  use the residue theorem straightlforwardly to obtain  ${\rm Res}[\mathcal{L}^{-1}[\mathcal{H}_\infty^{\rm SP}(s)]_{s_+}]$ and $ {\rm Res}[\mathcal{L}^{-1}[\mathcal{H}_\infty^{\rm SP}(s)]_{s_-}]$, respectively, as follows
	\begin{subequations}{\label{SolGPpm}}
		\begin{eqnarray}
			{\rm Res}[\mathcal{L}^{-1}[\mathcal{H}_\infty^{\rm GP}(s)]_{s_+}] 
			&=& -\frac{i}{2\omega} \frac{ e^{\frac{\tau^2(\gamma - i\omega)^2}{2}}e^{-t (\gamma -i \omega)}}{ \left(1-e^{T (\gamma - i \omega )}\right)},  \\
			{\rm Res}[\mathcal{L}^{-1}[\mathcal{H}_\infty^{\rm GP}(s)]_{s_-}]   
			&=&\frac{i}{2\omega} \frac{e^{\frac{\tau^2(\gamma +i\omega)^2}{2}} e^{-t (\gamma +i \omega )} }{
				\left(1-e^{T (\gamma +i \omega )}\right)}.
		\end{eqnarray}
	\end{subequations}
	Equation~(\ref{SolGPpm}) can be added and further simplified to be ${\rm Res}[\mathcal{L}^{-1}[\mathcal{H}_\infty^{\rm GP}(s)]_{s_{\pm}}]$ as
	\begin{equation}\label{GPxolp2}
		{\rm Res}[\mathcal{L}^{-1}[\mathcal{H}_\infty^{\rm GP}(s)]_{s_{\pm}}] =	e^{-\gamma t} e^{-\frac{1}{2}\tau^2(\omega^2-\gamma^2)} \, \frac{\sin(\omega(t-\gamma \tau^2)-e^{\gamma T}\sin(\omega(t+T-\gamma \tau^2)))}{\omega(1+e^{2\gamma T}-2e^{\gamma T}\cos(\omega T))} .
	\end{equation}
	
	Now, the infinite number of simple poles those appear only in the imaginary axis of the last term of Eq.~(\ref{GGPsa}), i.e., $s_k = i\omega_k = ik\omega_{\rm R}, k=\pm1, \pm2, \cdots,$ can be written in two separated terms corresponding for $k>0$, and $k <0$, repectively, as
	\begin{equation} 	
		\sum_{k = -\infty,\,k\ne0}^\infty  {\rm Res}[\mathcal{L}^{-1}[\mathcal{H}_\infty^{\rm GP}(s)]_{s_k}] = \sum_{k = 1}^{\infty}  {\rm Res}[\mathcal{L}^{-1}[\mathcal{H}_\infty^{\rm GP}(s)]_{s_k}]+\sum_{k = -1}^{-\infty}  {\rm Res}[\mathcal{L}^{-1}[\mathcal{H}_\infty^{\rm GP}(s)]_{s_k}],
	\end{equation}
	where $  \sum_{k = 1}^\infty  {\rm Res}[\mathcal{L}^{-1}[\mathcal{H}_\infty^{\rm GP}(s)]_{s_k}]$ and $ \sum_{k = -1}^{-\infty} {\rm Res}[\mathcal{L}^{-1}[\mathcal{H}_\infty^{\rm GP}(s)]_{s_{k}}]$ can be obtained from the Bromwich integral and the residue theorems as follows
	\begin{subequations}{\label{XsolpGP}}
		\begin{eqnarray}
			\sum_{k = 1}^\infty  {\rm Res}[\mathcal{L}^{-1}[\mathcal{H}_\infty^{\rm GP}(s)]_{s_k}]   
			&=& \frac{1}{T} \sum_{k=1}^\infty \frac{e^{-\frac{\tau^2\omega_k^2}{2}}e^{i\omega_k t}}{-\omega_k^2+2\gamma i \omega_k + \gamma^2+\omega^2}, \label{XsolGPp}\\
			\sum_{k = -1}^{-\infty}  {\rm Res}[\mathcal{L}^{-1}[\mathcal{H}_\infty^{\rm GP}(s)]_{s_{k}}]   
			&=& \frac{1}{T}\sum_{k=1}^\infty \frac{e^{-\frac{\tau^2\omega_k^2}{2}}e^{-i\omega_k t}}{-\omega_k^2-2\gamma i \omega_k + \gamma^2+\omega^2}.\label{XsolGPm}
		\end{eqnarray}
	\end{subequations}
	Thus, the two terms of Eqs.~(\ref{XsolGPp}) and (\ref{XsolGPm}) can be summed in a compact form as
	\begin{equation}\label{GPsum2}
		\sum_{k={-\infty},k\ne0}^\infty {\rm Res}[\mathcal{L}_{\rm p}^{\rm GP}(s)]_{s_k} =\frac{2}{T} \sum_{k=1}^\infty e^{-\frac{1}{2} \tau ^2 \omega _k^2} \, \frac{2 \gamma  \omega _k
			\sin \left(\omega _k t\right)+\left(\gamma ^2+\omega^2-\omega _k^2\right) \cos \left( \omega _k t\right)}{(\gamma^2+\omega^2-\omega_k^2)^2 +4\gamma^2\omega_k^2}.
	\end{equation}
	Therefore, we can find the analytical expression of the particular solution $x_{\rm p}^{\rm GP}(t)$ from Eqs.~(\ref{GPRes0}), (\ref{GPxolp2}), and (\ref{GPsum2}) as
	\begin{eqnarray}\label{GPpsolGP}
		x_{\rm \infty}^{\rm GP}(t) &=&   \frac{I_p}{mT}\frac{1}{\gamma^2+\omega^2}  + \frac{I_p}{m}e^{-\gamma t} e^{\frac{1}{2}\tau^2(\gamma^2-\omega^2)} \, \frac{\sin(\omega(t-\gamma \tau^2)-e^{\gamma T}\sin(\omega(t+T-\gamma \tau^2)))}{\omega(1+e^{2\gamma T}-2e^{\gamma T}\cos(\omega T))} \nonumber \\
		&&+\frac{2I_p}{mT}\sum_{k=1}^\infty e^{-\frac{1}{2} \tau ^2 \omega _k^2} \, \frac{2 \gamma  \omega _k
			\sin \left(\omega _k t\right)+\left(\gamma ^2+\omega^2-\omega _k^2\right) \cos \left( \omega _k t\right)}{
			(\gamma^2+\omega^2-\omega_k^2)^2 +4\gamma^2\omega_k^2}.
	\end{eqnarray}
	Finally, from Eq.~(\ref{AxpsolGP}), the particular solution driven by a train of Gaussian pulses could be written as 
	\begin{equation}\label{FinalGP}
		x_{\rm p}^{\rm GP}(t,T,\tau,Q) = x_{\infty}^{\rm GP}(t-Q)\Theta(t-Q),
	\end{equation}
	where $x_{\infty}^{\rm GP}(t)$ is given in Eq.~(\ref{GPpsolGP}). By taking the limit as $\tau\rightarrow0$ of Eq.~(\ref{FinalGP}), we immediately see that Eq.~(\ref{FinalGP}) becomes exactly the same as Eq.~(\ref{xpt}) as it should be.


\end{document}